\documentclass[aps,prc,showpacs,twocolumn,groupedaddress]{revtex4}

\usepackage{graphicx}
\usepackage{dcolumn}
\usepackage{psfrag,color}
\usepackage{subfigure}
\usepackage{amsfonts, amssymb, latexsym}
\usepackage{times}
\newcommand{\tc}{,~}

\newcommand{\moeller}{M$\o$ller }

\newcommand{\ket}[1]{\left| #1 \right\rangle}

\newcommand{\ua}{\uparrow}
\newcommand{\da}{\downarrow}

\newcommand{\Ua}{\Uparrow}
\newcommand{\Da}{\Downarrow}
\newcommand{\Ra}{\Rightarrow}
\newcommand{\La}{\Leftarrow}
\newcommand{\q}{\mathrm}

\newcommand{\duratio}{\mathit{R^{du}}}

\def\gtorder{\mathrel{\raise.3ex\hbox{$>$}\mkern-14mu
 \lower0.7ex\hbox{$\sim$}}}
\def\ltorder{\mathrel{\raise.3ex\hbox{$<$}\mkern-14mu
 \lower0.7ex\hbox{$\sim$}}}

\newcommand{\ncf}{1}
\newcommand{\ncaltech}{2}
\newcommand{\ncalsu}{3}
\newcommand{\nfiu}{4}
\newcommand{\nfsu}{5}
\newcommand{\nuiuc}{6}
\newcommand{\ninfn}{7}
\newcommand{\njlab}{8}
\newcommand{\nksu}{9}
\newcommand{\nkentucky}{10}
\newcommand{\numd}{11}
\newcommand{\numass}{12}
\newcommand{\nmit}{13}
\newcommand{\nunh}{14}
\newcommand{\nodu}{15}
\newcommand{\nrutgers}{16}
\newcommand{\nsaclay}{17}
\newcommand{\nsyracuse}{18}
\newcommand{\ntelaviv}{19}
\newcommand{\ntemple}{20}
\newcommand{\nuva}{21}
\newcommand{\nwm}{22}

\begin{document}

\preprint{APS/123-QED}


\title{Precision Measurement of the Neutron Spin 
Asymmetries and Spin-dependent Structure Functions 
in the Valence Quark Region}

\author{
X.~Zheng,$^{\nmit}$
K.~Aniol,$^{\ncalsu}$
D.~S.~Armstrong,$^{\nwm}$
T.~D.~Averett,$^{{\njlab,\nwm}}$
W.~Bertozzi,$^{\nmit}$
S.~Binet,$^{\nuva}$
E.~Burtin,$^{\nsaclay}$
E.~Busato,$^{\nrutgers}$ 
C.~Butuceanu,$^{\nwm}$
J.~Calarco,$^{\nunh}$
A.~Camsonne,$^{\ncf}$
G.~D.~Cates,$^{\nuva}$
Z.~Chai,$^{\nmit}$
J.-P.~Chen,$^{\njlab}$
Seonho~Choi,$^{\ntemple}$
E.~Chudakov,$^{\njlab}$
F.~Cusanno,$^{\ninfn}$
R.~De~Leo,$^{\ninfn}$
A.~Deur,$^{\nuva}$
S.~Dieterich,$^{\nrutgers}$ 
D.~Dutta,$^{\nmit}$
J.~M.~Finn,$^{\nwm}$
S.~Frullani,$^{\ninfn}$
H.~Gao,$^{\nmit}$ 
J.~Gao,$^{\ncaltech}$
F.~Garibaldi,$^{\ninfn}$
S.~Gilad,$^{\nmit}$
R.~Gilman,$^{{\njlab,\nrutgers}}$ 
J.~Gomez,$^{\njlab}$
J.-O.~Hansen,$^{\njlab}$
D.~W.~Higinbotham,$^{\nmit}$ 
W.~Hinton,$^{\nodu}$
T.~Horn,$^{\numd}$
C.W.~de~Jager,$^{\njlab}$
X.~Jiang,$^{\nrutgers}$ 
L.~Kaufman,$^{\numass}$ 
J.~Kelly,$^{\numd}$ 
W.~Korsch,$^{\nkentucky}$
K.~Kramer,$^{\nwm}$
J.~LeRose,$^{\njlab}$
D.~Lhuillier,$^{\nsaclay}$
N.~Liyanage,$^{\njlab}$
D.J.~Margaziotis,$^{\ncalsu}$
F.~Marie,$^{\nsaclay}$
P.~Markowitz,$^{\nfiu}$
K.~McCormick,$^{\nksu}$
Z.-E.~Meziani,$^{\ntemple}$
R.~Michaels,$^{\njlab}$
B.~Moffit,$^{\nwm}$
S.~Nanda,$^{\njlab}$
D.~Neyret,$^{\nsaclay}$
S.~K.~Phillips,$^{\nwm}$ 
A.~Powell,$^{\nwm}$
T.~Pussieux,$^{\nsaclay}$
B.~Reitz,$^{\njlab}$
J.~Roche,$^{\nwm}$ 
R.~Roch\'{e},$^{\nfsu}$
M.~Roedelbronn,$^{\nuiuc}$
G.~Ron,$^{\ntelaviv}$
M.~Rvachev,$^{\nmit}$ 
A.~Saha,$^{\njlab}$
N.~Savvinov,$^{\numd}$
J.~Singh,$^{\nuva}$
S.~\v{S}irca,$^{\nmit}$
K.~Slifer,$^{\ntemple}$
P.~Solvignon,$^{\ntemple}$
P.~Souder,$^{\nsyracuse}$
D.J.~Steiner,$^{\nwm}$ 
S.~Strauch,$^{\nrutgers}$
V.~Sulkosky,$^{\nwm}$ 
A.~Tobias,$^{\nuva}$
G.~Urciuoli,$^{\ninfn}$
A.~Vacheret,$^{\numass}$
B.~Wojtsekhowski,$^{\njlab}$
H.~Xiang,$^{\nmit}$ 
Y.~Xiao,$^{\nmit}$ 
F.~Xiong,$^{\nmit}$ 
B.~Zhang,$^{\nmit}$ 
L.~Zhu,$^{\nmit}$
X.~Zhu,$^{\nwm}$
P.A.~\.Zo{\l}nierczuk,$^{\nkentucky}$
}
\address{
\baselineskip 2 pt
\vskip 0.3 cm
{\rm The Jefferson Lab Hall A Collaboration} \break
\vskip 0.1 cm
{\small{
{$^{\ncf}$Universit\'{e} Blaise Pascal Clermont-Ferrand et CNRS/IN2P3 LPC 63\tc 177 Aubi\`{e}re Cedex\tc France} \break
{$^{\ncaltech}$California Institute of Technology\tc Pasadena\tc California 91125\tc USA} \break
{$^{\ncalsu}$California State University\tc Los Angeles\tc Los Angeles\tc California 90032\tc USA} \break
{$^{\nfiu}$Florida International University\tc Miami\tc Florida 33199\tc USA} \break
{$^{\nfsu}$Florida State University\tc Tallahassee\tc Florida 32306\tc USA} \break
{$^{\nuiuc}$University of Illinois\tc Urbana\tc Illinois 61801\tc USA} \break
{$^{\ninfn}$Istituto Nazionale di Fisica Nucleare\tc Sezione Sanit\`a\tc 00161 Roma\tc Italy} \break
{$^{\njlab}$Thomas Jefferson National Accelerator Facility\tc Newport News\tc Virginia 23606\tc USA} \break
{$^{\nksu}$Kent State University\tc Kent\tc Ohio 44242\tc USA} \break
{$^{\nkentucky}$University of Kentucky\tc Lexington\tc Kentucky 40506\tc USA} \break
{$^{\numd}$University of Maryland\tc College Park\tc Maryland 20742\tc USA} \break
{$^{\numass}$University of Massachusetts Amherst\tc Amherst\tc Massachusetts 01003\tc USA} \break
{$^{\nmit}$Massachusetts Institute of Technology\tc Cambridge\tc Massachusetts 02139\tc USA} \break
{$^{\nunh}$University of New Hampshire\tc Durham\tc New Hampshire 03824\tc USA} \break
{$^{\nodu}$Old Dominion University\tc Norfolk\tc Virginia 23529\tc USA} \break
{$^{\nrutgers}$Rutgers\tc The State University of New Jersey\tc Piscataway\tc New Jersy 08855\tc USA} \break
{$^{\nsaclay}$CEA Saclay\tc DAPNIA/SPhN\tc F-91191 Gif sur Yvette\tc France} \break
{$^{\nsyracuse}$Syracuse University\tc Syracuse\tc New York 13244\tc USA} \break
{$^{\ntelaviv}$University of Tel Aviv\tc Tel Aviv 69978\tc Israel} \break
{$^{\ntemple}$Temple University\tc Philadelphia\tc Pennsylvania 19122\tc USA} \break
{$^{\nuva}$University of Virginia\tc Charlottesville\tc Virginia 22904\tc USA} \break
{$^{\nwm}$College of William and Mary\tc Williamsburg\tc Virginia 23187\tc USA} \break
}}
}
\begin{abstract}                
We report on measurements of the neutron spin asymmetries $A_{1,2}^n$ and polarized 
structure functions $g_{1,2}^n$ at three kinematics in the deep inelastic region,
with $x=0.33$, $0.47$ and $0.60$ and $Q^2=2.7$, $3.5$ and $4.8$~(GeV/c)$^2$, respectively. 
These measurements were performed using a $5.7$ GeV longitudinally-polarized electron beam 
and a polarized $^3$He target. 
The results for $A_1^n$ and $g_1^n$ at $x=0.33$ are consistent with previous world data and, 
at the two higher $x$ points, have improved the precision of the world data by about 
an order of magnitude. The new $A_1^n$ data show a zero crossing around $x=0.47$ and 
the value at $x=0.60$ is significantly positive. These results agree with a 
next-to-leading order QCD analysis of previous world data. The trend of data at high 
$x$ agrees with constituent quark model predictions but disagrees with that from 
leading-order perturbative QCD (pQCD) assuming hadron helicity conservation.  
Results for $A_2^n$ and $g_2^n$ have a precision comparable to the best world data 
in this kinematic region.
Combined with previous world data, the moment $d_2^n$ was evaluated and the new 
result has improved the precision of this quantity by about a factor of two.
When combined with the world proton data, polarized quark distribution functions were 
extracted from the new $g_1^n/F_1^n$ values based on the quark parton model. 
While results for $\Delta u/u$ agree well 
with predictions from various models, results for $\Delta d/d$ disagree with the 
leading-order pQCD prediction when hadron helicity conservation is imposed. 
\end{abstract}

\maketitle

\section{INTRODUCTION}\label{ch1:main}

Interest in the spin structure of the nucleon became prominent 
in the 1980's when experiments at CERN~\cite{exp:cern-emc} 
and SLAC~\cite{exp:e080e130} on the integral of the proton polarized 
structure function $g_1^p$ showed that the total spin carried by quarks 
was very small, $\approx (12\pm 17)\%$~\cite{exp:cern-emc}. 
This was in contrast to the simple relativistic valence quark model 
prediction~\cite{theory:rCQM} in which the spin of the valence 
quarks carries approximately $75\%$ of the proton spin and the 
remaining $25\%$ comes from their orbital angular momentum. 
Because the quark model is very successful in describing static 
properties of hadrons, the fact that the quark spins account for only 
a small part of the nucleon spin was a big surprise and generated 
very productive experimental and theoretical activities to the present. 
Current understanding~\cite{theory:spin-sr} of the 
nucleon spin is that the total spin is distributed among valence 
quarks, $q\bar{q}$ sea quarks, their orbital angular momenta, and 
gluons. This is called the nucleon spin sum rule:
\begin{eqnarray}
 S_z^N &=& S_z^q+L_z^q+J_z^g = \frac{1}{2}~, \label{equ:spin}
\end{eqnarray}
where $S_z^N$ is the nucleon spin, $S_z^q$ and $L_z^q$ represent 
respectively the quark spin and orbital angular momentum (OAM), 
and $J_z^g$ is the total angular momentum of the gluons.
Only about $(20-30)\%$ of the nucleon spin is carried by the 
spin of the quarks. 
To further study the nucleon spin, one thus needs to know more
precisely how it decomposes into the three components and to measure
their dependence on $x$.
Here $x$ is the Bjorken scaling variable, which in the quark-parton 
model~\cite{theory:partonmodel} can be interpreted as the fraction 
of the nucleon momentum carried by the quark.  For a fixed target 
experiment one has $x={Q^2}/({2M\nu})$, with $M$ the nucleon mass, 
$Q^2$ the four momentum transfer squared and $\nu$ the energy transfer 
from the incident electron to the target.
However, due to experimental limitations, precision data have been 
collected so far only in the low and moderate $x$ regions. In these 
regions, one is sensitive to contributions from a large 
amount of $q\bar q$ sea and gluons and the nucleon is hard to model.
Moreover, at large
distances corresponding to the size of a nucleon, the theory of 
the strong interaction -- Quantum Chromodynamics (QCD) -- is highly 
non-perturbative, which makes the investigation of the roles of quark 
orbital angular momentum (OAM) and gluons in the nucleon spin 
structure difficult.

Our focus here is the first precise neutron spin structure data in the 
large $x$ region $x\gtorder 0.4$. For these kinematics, the valence quarks 
dominate and the ratios of structure functions can be estimated based on our 
knowledge of the interactions between quarks. More specifically, the virtual
photon asymmetry $A_1$, defined as
\begin{eqnarray}
 A_1(x,Q^2) &\equiv& \frac{\sigma_{{1}/{2}}-\sigma_{{3}/{2}}}
 {\sigma_{{1}/{2}}+\sigma_{{3}/{2}}}  \nonumber 
\end{eqnarray}
(the definitions of $\sigma_{1/2,3/2}$ are given in Appendix~\ref{app:formalism}),
which at large $Q^2$ is approximately 
the ratio of the polarized and the unpolarized structure functions 
$g_1/F_1$, is expected to approach unity as $x\to 1$ in 
perturbative QCD (pQCD). This is a 
dramatic prediction, not only because this is the only kinematic 
region where one can give an absolute prediction for the structure 
functions based on pQCD, but also because all previous data on 
the neutron asymmetry $A_1^n$ in the region $x\gtorder 0.4$
have large uncertainties and are consistent with $A_1^n\leqslant 0$.
Furthermore, because both $q\bar q$ sea and gluon contributions are 
small in this region, it is a relatively clean region to test the 
valence quark model and to study the role of 
valence quarks and their OAM contribution to the nucleon spin.\\

\smallskip
Deep inelastic scattering (DIS) has served as one of the 
major experimental tools to study the quark and gluon 
structure of the nucleon. The formalism of unpolarized and polarized
DIS is summarized in Appendix~\ref{app:formalism}.
Within the quark parton model (QPM), 
the nucleon is viewed as a collection of non-interacting, 
point-like constituents, one of which carries a fraction $x$ 
of the nucleon's longitudinal momentum and absorbs the 
virtual photon~\cite{theory:partonmodel}. The nucleon cross section 
is then the incoherent sum of the cross sections for elastic scattering 
from individual charged point-like partons.  Therefore the 
unpolarized and the polarized structure functions $F_1$
and $g_1$ can be related to the spin-averaged and spin-dependent 
quark distributions as~\cite{book:thomas&weise}
\begin{eqnarray}
 F_1(x,Q^2) =&&\hspace*{-0.4cm} \frac{1}{2}\sum_i e_i^2 q_i(x,Q^2)
 \label{equ:F1qpm}
\end{eqnarray}
and
\begin{eqnarray}
 g_1(x,Q^2) =&&\hspace*{-0.4cm} \frac{1}{2}\sum_i e_i^2 \Delta q_i(x,Q^2)~,
 \label{equ:g1qpm}
\end{eqnarray}
where $q_i(x,Q^2)=q_i^{\uparrow}(x,Q^2)+q_i^{\downarrow}(x,Q^2)$ is the 
unpolarized parton distribution function (PDF) of the $i^{th}$ 
quark, defined as the probability that the $i^{th}$ quark 
inside a nucleon carries a fraction $x$ of the nucleon's 
momentum, when probed with a resolution determined by $Q^2$.
The polarized PDF is defined as
$\Delta q_i(x,Q^2)=q_i^{\uparrow}(x,Q^2)-q_i^{\downarrow}(x,Q^2)$, 
where $q_i^\uparrow(x,Q^2)$ ($q_i^\downarrow(x,Q^2)$) is the probability 
to find the spin of the $i^{th}$ quark aligned parallel 
(anti-parallel) to the nucleon spin.

The polarized structure function $g_2(x,Q^2)$ does not have a simple 
interpretation within the QPM~\cite{book:thomas&weise}.  
However, it can be separated into leading twist and higher twist terms
using the operator expansion method~\cite{theory:ope}:
\begin{eqnarray} 
 g_2(x,Q^2) &=& g_2^{WW}(x,Q^2) +\bar g_2(x,Q^2)~.
\end{eqnarray} 
Here $g_2^{WW}(x,Q^2)$ is the leading twist (twist-2) contribution and can
be calculated using the twist-2 component of $g_1(x,Q^2)$ and the 
Wandzura-Wilczek relation~\cite{theory:g2ww} as
\begin{eqnarray} 
  g_2^{WW}(x,Q^2) &=& -g_1(x,Q^2) + \int_x^1
   \frac{g_1(y,Q^2)}{y}\mathrm{d}y ~. \label{equ:g2ww} 
\end{eqnarray}
The higher-twist contribution to $g_2$ is given by $\bar g_2$. 
When neglecting quark mass
effects, the higher-twist term represents interactions beyond the QPM,
{\it e.g.}, quark-gluon and quark-quark correlations~\cite{theory:ope-g2}.
The moment of $\bar g_2$ can be related to the matrix element 
$d_2$~\cite{theory:d2def}:
\begin{eqnarray}
  d_2 &=& \int_{0}^{1} \q{d}x ~x^2 \Big[3g_2(x,Q^2)+2g_1(x,Q^2)\Big] \nonumber\\
      &=& 3\int_{0}^{1} \q{d}x ~x^2 \bar g_2(x,Q^2)~. \label{equ:d2def}
\end{eqnarray}
Hence $d_2$ measures the deviations of $g_2$ from $g_2^{WW}$.
The value of $d_2$ can be obtained from measurements of 
$g_1$ and $g_2$ and can be compared with predictions from 
Lattice QCD~\cite{theory:d2lattice}, bag models~\cite{theory:d2bag},
QCD sum rules~\cite{theory:d2QCDSR} and chiral soliton 
models~\cite{theory:d2chi}.

\bigskip
In this paper we first describe 
available predictions for $A_1^n$ at large $x$.  
The experimental apparatus and the data 
analysis procedure will be described 
in Section~\ref{exp:main}, \ref{targ:main} and~\ref{ch5:main}.
In Section~\ref{result:main} we present
results for the asymmetries and polarized structure functions 
for both $^3$He and the neutron, a new experimental fit for 
$g_1^n/F_1^n$ and a result for the matrix element $d_2^n$.
Combined with the world proton and deuteron data, polarized quark 
distribution functions were extracted from our $g_1^n/F_1^n$ results. 
We conclude the paper by summarizing the results
for $A_1^n$ and $\Delta d/d$ and speculating on the importance of 
the role of quark OAM on the nucleon spin in the kinematic region 
explored.  
Some of the results presented here were published 
previously~\cite{A1nPRL};
the present publication gives full details on the experiment
and all of the neutron spin structure results for completeness.

\section{Predictions for $A_1^n$ at large $x$}\label{ch2:main}

From Section~\ref{ch2:su6} to \ref{ch2:other} we present
predictions of $A_1^n$ at large $x$. Data on $A_1^n$ from previous
experiments did not have the precision to distinguish 
among different predictions, as will be shown in Section~\ref{ch2:exp}.  

\subsection{SU(6) Symmetric Non-Relativistic Constituent Quark Model}
\label{ch2:su6}

In the simplest non-relativistic constituent quark 
model~(CQM)~\cite{theory:cqm_org}, the nucleon is made of three
constituent quarks and the nucleon spin is fully carried by the
quark spin.  Assuming SU(6) symmetry, the wavefunction of a neutron 
polarized in the $+z$ direction then has the form~\cite{theory:su6close}:
\begin{eqnarray}\label{equ:su6wvfunc}
  \ket{n\ua} = \frac{1}{\sqrt{2}}\ket{d^\ua (du)_{000}}
   +\frac{1}{\sqrt{18}}\ket{d^\ua (du)_{110}}~~~~~~~~~~~~~~~~~~&&  \\
     -\frac{1}{3} \ket{d^\da (du)_{111}}
     -\frac{1}{3}\ket{u^\ua (dd)_{110}}
     +\frac{\sqrt{2}}{3}\ket{u^\da (dd)_{111}},&&\nonumber
\end{eqnarray}
where the three subscripts are the total isospin, total spin $S$ 
and the spin projection $S_z$ along the $+z$ direction for the 
`diquark' state. For the case of a proton one needs to exchange 
the $u$ and $d$ quarks in Eq.~(\ref{equ:su6wvfunc}). 
In the limit where SU(6) symmetry is exact, both diquark spin states with 
$S=1$ and $S=0$ contribute equally to the observables of interest, 
leading to the predictions
\begin{eqnarray}
  && A_1^p={5}/{9}~~\q{and}~A_1^n=0 ~; \\
  &&\Delta u/u \to 2/3~~\q{and}~ \Delta d/d\to -1/3.
\end{eqnarray}

We define $u(x)\equiv u^p(x)$, $d(x)\equiv d^p(x)$ and 
$s(x)\equiv s^p(x)$ as parton distribution functions (PDF) for the 
proton.  For a neutron one has $u^n(x)=d^p(x)=d(x)$, $d^n(x)=u^p(x)=u(x)$ 
based on isospin symmetry.  The strange quark distribution for the 
neutron is assumed to be the same as that of the proton, $s^n(x)=s^p(x)=s(x)$.
In the following, all PDF's are for the proton, unless specified by a 
superscript `n'.

In the case of DIS, exact SU(6) symmetry implies the same shape for the 
valence quark distributions, {\it i.e.} $u(x)=2d(x)$.  
Using Eq.~(\ref{equ:F1qpm}) and~(\ref{equ:disF1F2}), and assuming that $R(x,Q^2)$
is the same for the neutron and the proton, one can write the 
ratio of neutron and proton $F_2$ structure functions as
\begin{eqnarray}
  R^{np}\equiv\frac{F_2^n}{F_2^p}
	 = \frac{u(x)+4d(x)}{4u(x)+d(x)} ~.
\end{eqnarray}
\noindent Applying $u(x)=2d(x)$ gives
\begin{eqnarray}
  R^{np} = {2}/{3} ~.
\end{eqnarray}
However, data on the $R^{np}$ ratio from SLAC~\cite{data:slacRnp},
CERN~\cite{data:bcdmsRnp,data:emcRnp,data:nmcRnp} and Fermilab~\cite{data:e665Rnp}
disagree with this SU(6) prediction. The data show that $R^{np}(x)$ is a 
straight line starting with $R^{np}\vert_{x\to 0}\approx 1$ and 
dropping to below $1/2$ as $x\to 1$.  In addition, $A_1^p(x)$ is small at 
low $x$~\cite{data:a1pg1p-emc,data:a1pg1p-smc,data:a1pa1n-e143}. The fact 
that $R^{np}\vert_{x\to 0}\approx 1$ may be explained by the presence of 
a dominant amount of sea quarks in the low $x$ region and the fact that 
$A_1^{p}\vert_{x\to 0}\approx 0$ could be because these sea quarks are 
not highly polarized.  At large~$x$, however, there are few sea quarks 
and the deviation from SU(6) prediction indicates a problem with the 
wavefunction described by Eq.~(\ref{equ:su6wvfunc}). In fact, SU(6) 
symmetry is known to be broken~\cite{theory:su6breaking} and the details 
of possible SU(6)-breaking mechanisms is an important open issue in 
hadronic physics.

\subsection{SU(6) Breaking and Hyperfine Perturbed Relativistic CQM}
\label{ch2:cqm}

A possible explanation for the SU(6) symmetry breaking is the one-gluon exchange 
interaction which dominates the quark-quark interaction at short-distances. 
This interaction was used to explain the behavior of $R^{np}$ near 
$x\to 1$ and the $\approx 300$-MeV mass shift between the nucleon and the 
$\Delta(1232)$~\cite{theory:su6breaking}.
Later this was described by an interaction term proportional to 
$\vec{S}_i\cdot\vec{S}_j~\delta^3(\vec{r}_{ij})$, with $\vec{S_i}$ 
the spin of the $i^{th}$ quark, hence is also called the hyperfine 
interaction, or chromomagnetic interaction among the 
quarks~\cite{theory:hyperfine-spin}. The effect of this perturbation on 
the wavefunction is to lower the energy of the $S=0$ diquark state, 
causing the first term of Eq.~(\ref{equ:su6wvfunc}), 
$\ket{d\ua (ud)_{000}}^\q{n}$, 
to become more stable and to dominate the high energy tail of the quark 
momentum distribution that is probed as $x\to 1$.  Since the struck 
quark in this term has its spin parallel to that of the nucleon, the 
dominance of this term as $x\to 1$ implies $(\Delta d/d)^\q{n}\to 1$
and $(\Delta u/u)^\q{n}\to -1/3$ for the neutron, while for the proton 
one has
\begin{eqnarray}
 &&\Delta u/u \to 1~\q{and}~ \Delta d/d\to -1/3 \mathrm{~~as~} x\to 1 ~.
\end{eqnarray}
One also obtains
\begin{eqnarray}
 && R^{np}\to {1}/{4}~~\q{as}~ x\to 1 ~,
\end{eqnarray}
which could explain the deviation of $R^{np}(x)$ data from the 
SU(6) prediction. Based on the same mechanism, one can make the 
following predictions:
\begin{eqnarray}
  A_1^{p}\to 1~\q{and}~ A_1^{n}\to 1 ~~\q{as}~ x\to 1~.
\end{eqnarray}

The hyperfine interaction is often used to break SU(6) symmetry in 
the relativistic CQM (RCQM). In this model, the constituent quarks have 
non-zero OAM which carries $\approx 25\%$ of the nucleon 
spin~\cite{theory:rCQM}. The use of RCQM to predict the large $x$ 
behavior of the nucleon structure functions can be justified by the 
valence quark dominance, {\it i.e.}, in the large~$x$ region almost 
all quantum numbers, momentum and the spin of the nucleon are carried 
by the three valence quarks, which can therefore be identified as 
constituent quarks. 
Predictions of $A_1^n$ and $A_1^p$ in the large $x$ region using
the hyperfine-perturbed RCQM have been achieved~\cite{theory:cqm}.

\subsection{Perturbative QCD and Hadron Helicity Conservation}
\label{ch2:pqcd}

In the early 1970's, in one of the first applications of 
perturbative QCD (pQCD), it was noted that as $x\to 1$, the 
scattering is from a high-energy quark and thus the process can be 
treated perturbatively~\cite{theory:farrar}. 
Furthermore,
when the quark OAM is assumed to be zero, the conservation of angular 
momentum requires that a quark carrying nearly all the momentum of 
the nucleon ({\it i.e.} $x\to 1$) must have the same helicity as 
the nucleon. This mechanism is called hadron helicity 
conservation (HHC),
and is referred to as the leading-order pQCD in this paper. 
In this picture, quark-gluon interactions 
cause only the $S=1$, $S_z=1$ diquark spin projection component 
rather than the full $S=1$ diquark system to be suppressed as 
$x\to 1$, which gives
\begin{eqnarray}
 && \Delta u/u \to 1~\q{and}~ \Delta d/d\to 1 \q{~~as~} x\to 1~; \\
 && R^{np}\to \frac{3}{7},~
 A_1^{p}\to 1 \q{~~and~} A_1^{n}\to 1 \q{~~as~} x\to 1~.
\end{eqnarray}
This is one of the few places where pQCD can make an absolute 
prediction for the $x$-dependence of the structure functions 
or their ratios. However, how low in $x$ 
and $Q^2$ this picture works is uncertain.
HHC has been used as a constraint in a model to fit data
on the first moment of the proton $g_1^p$, 
giving the BBS parameterization~\cite{theory:bbs}. 
The $Q^2$ evolution was not included in this calculation.
Later in the LSS(BBS) parameterization~\cite{theory:lssbbs},
both proton and neutron $A_1$ data were fitted 
directly and the $Q^2$ evolution was carefully treated.  
Predictions for $A_1^n$ using both BBS and LSS(BBS) 
parameterizations have been made, as shown in 
Fig.~\ref{fig:a1nmodel_all} and \ref{fig:a1pmodel_all} in Section~\ref{ch2:exp}.


HHC is based on the assumption that the quark OAM is zero. Recent 
experimental data on the tensor polarization in elastic $e-^2$H 
scattering~\cite{data:cebaf-t20}, neutral pion 
photo-production~\cite{data:cebaf-gammap} and the proton electro-magnetic
form factors~\cite{data:cebaf-F2p,data:cebaf-Gep} disagree with the 
HHC predictions~\cite{theory:F2pF1pQCDscaling}.  It has been suggested 
that effects beyond leading-order pQCD, such as quark 
OAM~\cite{theory:miller,theory:transpdf,theory:ji1,theory:ji2},
might play an important role in processes involving quark spin flips.

\subsection{Predictions from Next-to-Leading Order QCD Fits}

In a next-to-leading order (NLO) QCD analysis of the world 
data~\cite{theory:lss2001}, parameterizations of the polarized 
and unpolarized PDFs were performed without the HHC constraint. 
Predictions of $g_1^p/F_1^p$ and $g_1^n/F_1^n$ were made
using these parameterizations,
as shown in Fig.~\ref{fig:a1nmodel_all} and 
\ref{fig:a1pmodel_all} in Section~\ref{ch2:exp}.

In a statistical approach, the nucleon is viewed as a gas of 
massless partons (quarks, antiquarks and gluons) in equilibrium 
at a given temperature in a finite volume, and the parton 
distributions are parameterized using either Fermi-Dirac or 
Bose-Einstein distributions. Based on this statistical picture 
of the nucleon, a global NLO QCD analysis of unpolarized and 
polarized DIS data was performed~\cite{theory:stat}.  In this 
calculation $\Delta u/u\approx 0.75$, 
$\Delta d/d\approx -0.5$ and $A_1^{p,n}<1$ at $x\to 1$.

\subsection{Predictions from Chiral Soliton and Instanton Models}

While pQCD works well in high-energy hadronic physics, 
theories suitable for hadronic phenomena in the 
non-perturbative regime are much more difficult to construct.
Possible approaches in this regime are quark models, chiral 
effective theories and the lattice QCD method.  
Predictions for $A_1^{n,p}$ have been made using chiral 
soliton models~\cite{theory:chi_weigel,theory:chi_waka} and the 
results of Ref.~~\cite{theory:chi_waka} give $A_1^n<0$.
 The prediction that $A_1^{p}<0$ has also been made
in the instanton model~\cite{theory:instanton}.

\subsection{Other Predictions}\label{ch2:other}

Based on quark-hadron duality~\cite{theory:duality-b&g}, one can obtain
the structure functions and their ratios in the large $x$ region by summing 
over matrix elements for nucleon resonance transitions. To incorporate 
SU(6) breaking, different mechanisms consistent with duality were assumed 
and data on the structure function ratio $R^{np}$ were used to fit the 
SU(6) mixing parameters. In this picture, $A_1^{n,p}\to 1$ as $x\to 1$ 
is a direct result. Duality predictions for $A_1^{n,p}$ using different SU(6) 
breaking mechanisms were performed in Ref.~\cite{theory:dual_new}. There also 
exist predictions from bag models~\cite{theory:bag},
as shown in Fig.~\ref{fig:a1nmodel_all} and 
\ref{fig:a1pmodel_all} in the next section.

\subsection{Previous Measurements of $A_1^n$}\label{ch2:exp}

A summary of previous $A_1^n$ measurements is given in 
\begin{table}[!h]
\caption{Previous measurements of $A_1^n$.}\label{tab:exA1nList}
\begin{center}
\begin{ruledtabular}
\begin{tabular}{c|c|c|c|c}
  Experiment & beam & target & $x$       & $Q^2$     \\ 
             &      &        &           & (GeV/c)$^2$ \\ \hline
  E142 \cite{data:a1ng1n-e142} & 19.42, 22.66,                 &
                  $^3$He & 0.03-0.6 & 2                \\
                               &               25.51 GeV; e$^-$ &
                         &               &                  \\
  E154 \cite{data:a1ng1n-e154} & 48.3 GeV; e$^-$                &
                  $^3$He  & 0.014-0.7  & 1-17            \\
  HERMES \cite{data:a1ng1n-hermes} &  27.5 GeV; e$^+$         &
                  $^3$He  & 0.023-0.6  & 1-15            \\
  E143 \cite{data:a1pa1n-e143} & 9.7, 16.2,    &
                  NH$_3$, ND$_3$ & 0.024-0.75 & 0.5-10          \\
                               & 29.1 GeV; e$^-$     &
                                            &    &                    \\
  E155 \cite{data:g1pg1n-e155} &  48.35 GeV; e$^-$              &
                  NH$_3$, LiD$_3$ & 0.014-0.9  & 1-40             \\ 
  SMC \cite{data:g1n-smc} &   190 GeV; $\mu^-$    &
                  C$_4$H$_{10}$O  &0.003-0.7& 1-60           \\
                                   &                      &
                  C$_4$D$_{10}$O  &         &                          \\
  \end{tabular}
 \end{ruledtabular}
 \end{center}
\end{table}
\begin{figure}[!h]
 \vspace{0.1in}
 \includegraphics[angle = 0, width=240pt]{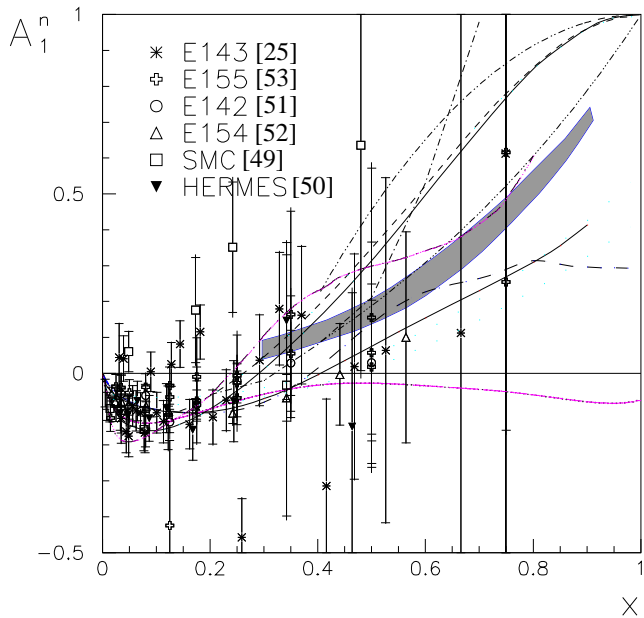}
 \caption{Previous data on $A_1^n$~\cite{data:a1ng1n-e142,
data:a1ng1n-e154,data:g1n-smc,data:a1ng1n-hermes,data:a1pa1n-e143,
data:g1pg1n-e155} and various theoretical predictions: 
$A_1^n$ from SU(6) symmetry (solid line at zero)~\cite{theory:su6close},
hyperfine-perturbed RCQM (shaded band)~\cite{theory:cqm},
BBS parameterization at $Q^2=4$~(GeV/c)$^2$ (higher solid)~\cite{theory:bbs},
LSS(BBS) parameterization at $Q^2=4$~(GeV/c)$^2$ (dashed)~\cite{theory:lssbbs},
statistical model at $Q^2=4$~(GeV/c)$^2$ (long-dashed)~\cite{theory:stat},
quark-hadron duality using two different SU(6) breaking mechanisms
(dash-dot-dotted and dash-dot-dot-dotted)\cite{theory:dual_new},
and non-meson cloudy bag model (dash-dotted)~\cite{theory:bag};
$g_1^n/F_1^n$ from LSS2001 parameterization at $Q^2=5$~(GeV/c)$^2$
(lower solid)~\cite{theory:lss2001}
and from chiral soliton 
models~\cite{theory:chi_weigel} at $Q^2=3$~(GeV/c)$^2$ (long dash-dotted)
and~\cite{theory:chi_waka} at $Q^2=4.8$~(GeV/c)$^2$ (dotted).
}
 \label{fig:a1nmodel_all}
\end{figure}
\begin{figure}[!h]
 \vspace{0.1in}
 \includegraphics[angle = 0, width=240pt]{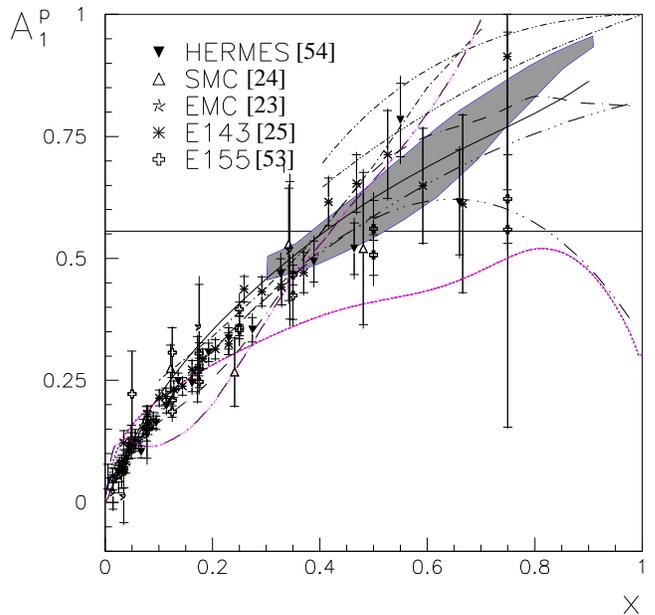}
 \caption{World data on 
$A_1^p$~\cite{data:a1pa1n-e143,data:a1pg1p-emc,data:a1pg1p-smc,
data:g1pg1n-e155,data:g1p-hermes} and 
predictions for $g_1^p/F_1^p$ at $Q^2=5$~(GeV/c)$^2$ from
the E155 experimental fit (long dash-dot-dotted)~\cite{data:g1pg1n-e155} 
and a new fit as described in Section~\ref{result:neutron} (long dash-dot-dot-dotted).
The solid curve corresponds to the prediction for $g_1^n/F_1^n$ from 
LSS(2001) parameterization at $Q^2=5$~(GeV/c)$^2$. Other curves are 
the same as in Fig.~\ref{fig:a1nmodel_all} except that there is no 
prediction for the proton from BBS and LSS(BBS) parameterizations.
}
 \label{fig:a1pmodel_all}
\end{figure}
Table~\ref{tab:exA1nList}.  The data on $A_1^n$ 
and $A_1^p$ are plotted in Fig.~\ref{fig:a1nmodel_all} 
and~\ref{fig:a1pmodel_all} along with theoretical calculations 
described in previous sections. Since the $Q^2$-dependence 
of $A_1$ is small and $g_1/F_1\approx A_1$ in DIS, data for $g_1^n/F_1^n$ 
and $g_1^p/F_1^p$ are also shown and all data are plotted without evolving
in $Q^2$. As becomes obvious in Fig.~\ref{fig:a1nmodel_all}, the precision of previous 
$A_1^n$ data at $x>0.4$ from SMC~\cite{data:g1n-smc}, 
HERMES~\cite{data:a1ng1n-hermes}
and SLAC~\cite{data:a1ng1n-e142,data:a1pa1n-e143,data:a1ng1n-e154} 
is not sufficient to distinguish among different
predictions.

\section{{THE EXPERIMENT}}\label{exp:main}

We report on an experiment~\cite{exp:e99117} carried out at in
the Hall~A of Thomas Jefferson National Accelerator Facility 
(Jefferson Lab, or JLab).
The goal of this experiment was to provide 
precise data on $A_1^n$ in the large $x$ region. We have measured
the inclusive deep inelastic scattering of longitudinally 
polarized electrons off a polarized $^3$He target, with the latter
being used as 
an effective polarized neutron target. The scattered electrons were 
detected by the two standard High Resolution Spectrometers (HRS). 
The two HRS were configured at the same scattering angles and momentum 
settings 
to double the statistics. Data were collected at three $x$ points as 
shown in Table~\ref{tab:kine}. 
Both longitudinal and transverse electron asymmetries were measured, from 
which $A_1$, $A_2$, $g_1/F_1$ and $g_2/F_1$ were extracted using 
Eq.~(\ref{equ:a2a1}--\ref{equ:a2g2}).
\begin{table}[ht]
\begin{center}
{\caption{Kinematics of the experiment. The beam energy was $E=5.734$~GeV.
$E^\prime$ and $\theta$ are the nominal momentum and angle of the scattered
electrons. $\langle x\rangle$, $\langle Q^2\rangle$ and $\langle W^2\rangle$ 
are values averaged over the spectrometer acceptance.}\label{tab:kine}}
\begin{ruledtabular}
\begin{tabular}{c|ccc}
  $\langle x\rangle$                & 0.327      & 0.466      & 0.601 \\ \hline
  $E^\prime$           & 1.32       & 1.72       & 1.455   \\
  $\theta$             & $35^\circ$ & $35^\circ$ & $45^\circ$  \\
  $\langle Q^2\rangle$ (GeV/c)$^2$  & 2.709      & 3.516      & 4.833 \\
  $\langle W^2\rangle$ (GeV)$^2$    & 6.462      & 4.908      & 4.090 \\
\end{tabular}
\end{ruledtabular}
\end{center}
\end{table}

\subsection{Polarized $^3$He as an Effective Polarized Neutron}\label{ch3:intro}

As shown in Fig.~\ref{fig:a1nmodel_all}, previous data on $A_1^n$ 
did not have sufficient precision in the large $x$ region. This is mainly
due to two experimental limitations. Firstly, high polarization and 
luminosity required for precision measurements in the large~$x$ region
were not available previously.
Secondly, there exists no free dense neutron target suitable for a 
scattering experiment, mainly because of the neutron's short lifetime 
($\approx 886$~sec).
Therefore polarized nuclear 
targets such as $^2\vec{\q{H}}$ or $^3\vec{\q{He}}$ are commonly used 
as effective polarized neutron targets.  Consequently, nuclear corrections 
need to be applied to extract neutron results from nuclear data.  
\begin{figure}[ht]
 \begin{center}
 \includegraphics[scale=0.66]{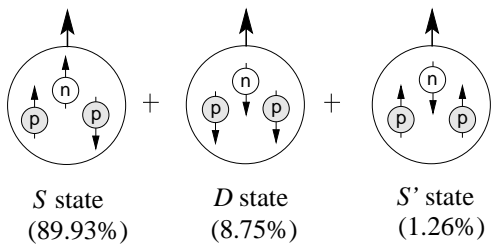}
 \end{center}
 \caption{An illustration of $^3$He wavefunction. The $S$, $S^\prime$ 
and $D$ state contributions are from calculations using the AV18 
two-nucleon interaction and the Tucson-Melbourne three-nucleon force, 
as given in Ref.~\cite{theory:PnPp_nogga}. 
}\label{fig:he3wavefunc}
\end{figure}
For a polarized deuteron, approximately half of the deuteron spin 
comes from the proton and the other half comes from the neutron.
Therefore the neutron results extracted from the deuteron data 
have a significant uncertainty coming from the error in the 
proton data.  
The advantage of using $^3\vec{\q{He}}$ is that the two protons' 
spins cancel in the dominant $S$ state of the $^3$He wavefunction, 
thus the spin of the $^3$He comes mainly ($>87\%$) from the 
neutron~\cite{theory:PnPp_friar,theory:PnPp_nogga}, as illustrated 
in Fig.~\ref{fig:he3wavefunc}. As a result, there is less model 
dependence in the procedure of extracting the spin-dependent 
observables of the neutron from $^3$He data. At large~$x$, the 
advantage of using a polarized $^3$He 
target is more prominent in the case of $A_1^n$.  In this region 
almost all calculations show that $A_1^n$ is much smaller than 
$A_1^p$, therefore the $A_1^n$ results
extracted from nuclear data are more sensitive to the uncertainty in the
proton data and the nuclear model being used.

In the large~$x$ region, the cross sections are small because the parton
densities drop dramatically as $x$ increases. In addition, the Mott 
cross section, given by Eq.~\ref{equ:Mottxsec}, is small at large $Q^2$.
To achieve a good statistical precision, high luminosity is required. 
Among all laboratories which are equipped with 
a polarized $^3$He target and are able to perform a measurement of 
the neutron spin structure, the polarized electron beam at JLab, combined 
with the polarized $^3$He target in Hall A, provides the highest polarized 
luminosity in the world~\cite{thesis:zheng}.  Hence it is the best place 
to study the large~$x$ behavior of the neutron spin structure.

\subsection{The Accelerator and the Polarized Electron Source}
\label{ch3:beam}

JLab operates a continuous-wave electron accelerator that 
recirculates the beam up to five times through two super-conducting linear 
accelerators. Polarized electrons are extracted from a strained 
GaAs photocathode~\cite{exp:polestrain} illuminated by circularly 
polarized light, providing a polarized beam of $(70-80)\%$ polarization 
and $\approx 200\mu A$ maximum current to experimental halls A, B and C.  
The maximum beam energy available at JLab so far is 5.7~GeV, which was
also the beam energy used during this experiment.  

\subsection{Hall~A Overview}
\begin{figure}[ht]
 \begin{center}
  \includegraphics[scale=0.45]{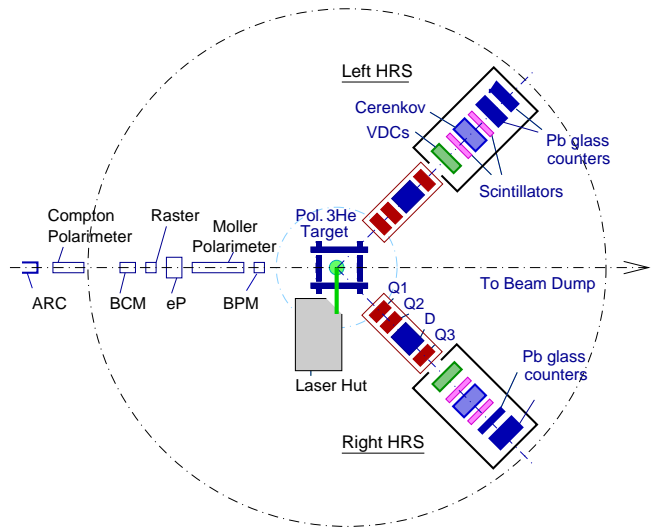}
 \end{center}
 \caption{(Color online) Top-view of the experimental hall A (not to scale).}\label{fig:floorplan}
\end{figure}

The basic layout of Hall~A during this experiment is shown in 
Fig.~\ref{fig:floorplan}.  The major instrumentation~\cite{exp:NIM} 
includes beamline equipment, the target and two HRSs.  
The beamline starts after the arc section of the accelerator
where the beam is bent into the hall, and ends at the beam dump.  
The arc section can be used for beam energy measurement, as will be 
described in Section~\ref{ch3:beamenergy}. After the arc section, 
the beamline is equipped with a Compton polarimeter, two Beam 
Current Monitors (BCM) and an Unser monitor for 
absolute beam current measurement, a fast raster, the eP device for 
beam energy measurement, a \moeller polarimeter and 
two Beam Position Monitors (BPM).  These beamline elements, together
with spectrometers and the target, will be 
described in detail in the following sections.

\subsection{Beam Energy Measurement}\label{ch3:beamenergy}
\indent
The energy of the beam was measured absolutely by two independent methods 
- ARC and eP~\cite{exp:NIM,exp:NIM-52}. Both methods can provide a 
precision of $\delta E_{beam}/E_{beam}\approx 2\times10^{-4}$.
For the ARC method~\cite{exp:NIM,thesis:marchand}, the deflection 
of the beam in the arc section of the beamline is used to determine 
the beam energy. In the eP measurement~\cite{exp:NIM,thesis:ravel}
the beam energy is determined by the measurement of the scattered 
electron angle $\theta_e$ and the recoil proton angle $\theta_p$ in 
$^1$H$(e,e'p)$ elastic scattering. 

\subsection{Beam Polarization Measurement} \label{ch3:beamPol}
Two methods were used during this experiment to measure the electron 
beam polarization.
%
The \moeller polarimeter~\cite{exp:NIM} measures \moeller scattering 
of the polarized electron beam off polarized atomic electrons in a 
magnetized foil. The cross section of this process depends on the 
beam and target polarizations. The polarized electron target used 
by the \moeller polarimeter was a ferromagnetic foil, with its 
polarization determined from foil magnetization measurements. The 
\moeller measurement is invasive and typically takes an hour, 
providing a statistical accuracy of about $0.2\%$. The systematic 
error comes mainly from the error in the foil target polarization.  
An additional systematic error is due to the fact that the beam 
current used during a \moeller measurement ($\approx 0.5\mu$A)
is lower than that used during the experiment. The total relative 
systematic error was $\approx 3.0\%$ during this experiment.

\smallskip
During a Compton polarimeter~\cite{exp:NIM,thesis:baylac} measurement, 
the electron beam is scattered off a circularly polarized photon beam 
and the counting rate asymmetry of the Compton scattered electrons 
or photons between opposite beam helicities is measured. The Compton 
polarimeter measures the beam polarization concurrently with the 
experiment running in the hall.

The Compton polarimeter
consists of a magnetic chicane which deflects the electron beam 
away from the scattered photons, a photon source, an electromagnetic 
calorimeter and an electron detector.  The photon source was a 
200~mW laser amplified by a resonant Fabry-Perot cavity.  During 
this experiment the maximum gain of the cavity reached $G_{max}=7500$, 
leading to a laser power of $1500$~W inside the cavity. The circular 
polarization of the laser beam was $>99\%$ for both right and left 
photon helicity states.  The asymmetry measured in Compton scattering 
at JLab with a $1.165$~eV photon beam and the $5.7$~GeV electron beam
used by this experiment had a mean value of $\approx 2.2\%$ and a 
maximum of $9.7\%$.  For a 12~$\mu$A beam current, one hour was 
needed to reach a relative statistical accuracy of 
$(\Delta P_b)_{stat}/P_b\approx 1\%$.
The total systematic error was $(\Delta P_b)_{sys}/P_b\approx 1.6\%$
during this experiment.

\smallskip
The average beam polarization during this experiment 
was extracted from a combined analysis of 7 \moeller and 53 Compton measurements.
A value of $(79.7\pm 2.4)\%$ was used in the final DIS analysis.

\subsection{Beam Helicity}\label{exp:helicity}

The helicity state of electrons is regulated every $33$~ms at the 
electron source. The time sequence of the electrons' helicity state is 
carried by helicity signals, which are sent to experimental halls and the 
data acquisition (DAQ) system. Since the status of the helicity signal 
(H+ or H- pulses) has either the same or the opposite sign as the real 
electron helicity, the absolute helicity state of the beam 
needs to be determined by other methods, as will be described later.

There are two modes -- toggle and pseudorandom -- which can be used 
for the pulse sequence of the helicity signal.  In the toggle mode, 
the helicity alternates every $33$~ms.  
In the pseudorandom mode, the helicity alternates randomly at the 
beginning of each pulse pair, of which the two pulses must have 
opposite helicities in order to equalize the numbers of the H+ and 
H- pulses.  The purpose of the pseudorandom mode is to minimize any 
possible time-dependent systematic errors. 
\begin{figure}[htp]
 \begin{center}
  \includegraphics[scale=0.32]{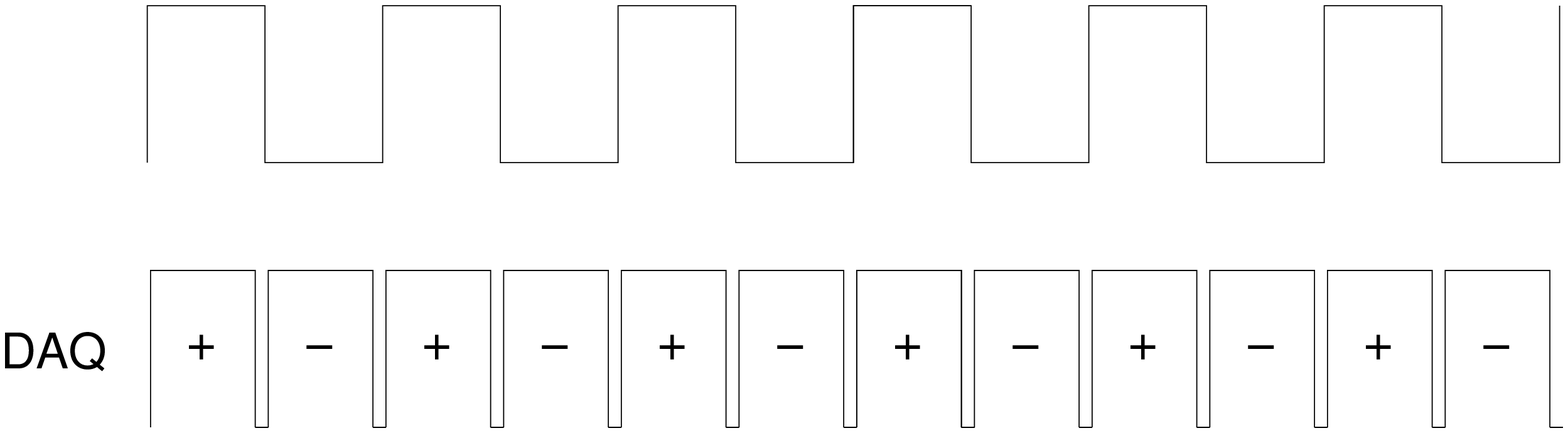}\\
  \vspace*{0.8cm}
  \includegraphics[scale=0.32]{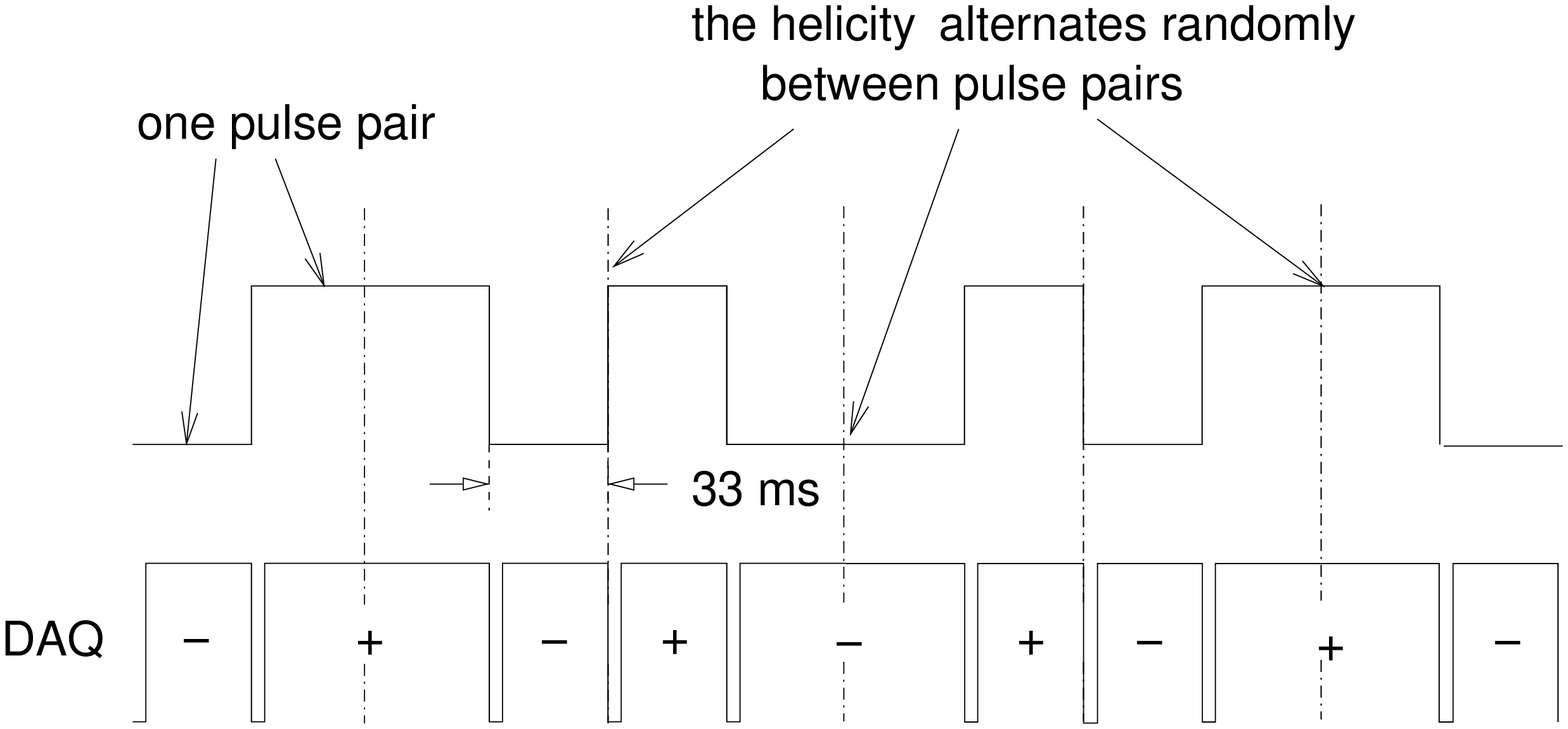}
 \end{center}
 \caption{Helicity signal and the helicity status of DAQ 
in toggle (top) and pseudorandom (bottom) modes.}
 \label{fig:helpulse}
\end{figure}
Fig.~\ref{fig:helpulse} shows the helicity signals and the 
helicity states of the DAQ system for the two regulation modes.

There is a half-wave plate at the polarized source which can be 
inserted to reverse the helicity of the laser illuminating the 
photocathode hence reverse the helicity of electron beam.  
During the experiment this half-wave plate was inserted for half 
of the statistics to minimize possible systematic effects related 
to the beam helicity.

The scheme described above was used to monitor the 
relative changes of the helicity state.  The absolute sign of 
the electrons' helicity states during each of the H+ and H- 
pulses were confirmed by measuring a well known asymmetry and 
comparing the measured asymmetry with its prediction, as will 
be presented in Section~\ref{ana:elana} and \ref{ana:delta}. 

\subsection{Beam Charge Measurement and Charge Asymmetry Feedback}\label{ch3:bcm}

The beam current was measured by the BCM system located upstream of the
target on the beamline.  The BCM signals were fed to scaler 
inputs and were inserted in the data stream.

Possible beam charge asymmetry measured at Hall~A can be caused by 
the timing asymmetry of the DAQ system, or by the timing and
the beam intensity asymmetries at the 
polarized electron source.  The beam intensity asymmetry originates from the 
intensity difference between different helicity states of the circularly 
polarized laser used to strike the photocathode. Although the charge
asymmetry can be corrected for to first order, there may exist 
unknown non-linear effects which can cause a systematic error in the 
measured asymmetry. Thus the beam charge asymmetry should be
minimized. This was done by using a separate DAQ system initially developed 
for the parity-violation experiments~\cite{exp:parityQasym},
called the parity DAQ.
The parity DAQ used the measured charge asymmetry in Hall~A to control
the orientation of a rotatable half-wave plate located before the 
photocathode at the source, such that intensities for each helicity state
of the polarized laser used to strike the photocathode were adjusted 
accordingly. The parity DAQ was synchronized with the two HRS DAQ 
systems so that the charge asymmetry in the two different helicity states 
could be monitored for each run.  The charge
asymmetry was typically controlled to be below $2\times 10^{-4}$ during this experiment.

\subsection{Raster and Beam Position Monitor}\label{ch3:bpm}

To protect the target cell from being damaged by the effect of beam-induced 
heating, the beam was rastered at the target. The raster consists a pair of 
horizontal and vertical air-core dipoles located upstream of the target on 
the beamline, which can produce either a rectangular or an elliptical pattern.
We used a raster pattern distributed uniformly over a circular area 
with a radius of 2~mm.

The position and the direction of the beam at the target were 
measured by two BPMs located 
upstream of the target~\cite{exp:NIM}. The beam position can be
measured with a precision of 200~$\mu$m with respect to the Hall~A coordinate 
system. The beam position and angle at the target were recorded for each event.

\subsection{High Resolution Spectrometers}\label{ch3:hrs}

The Hall~A High Resolution Spectrometer (HRS) systems were designed 
for detailed investigations of the structure of nuclei and nucleons. 
They provide high resolution in momentum and in angle reconstruction 
of the reaction product as well as being able to be operated at high 
luminosity.  For each spectrometer, the vertically bending 
design includes two quadrupoles followed by a dipole magnet and 
a third quadrupole. All quadrupoles and the dipole are superconducting. 
Both HRSs can provide a momentum resolution better than $2\times 10^{-4}$ 
and a horizontal angular resolution better than 2~mrad with a 
design maximum central momentum of 4~GeV/c~\cite{exp:NIM}.
By convention, the two spectrometers are identified as the left
and the right spectrometers based on their position when viewed
looking downstream.
\begin{figure}[htp]
 \begin{center}
  \includegraphics[scale=0.6]{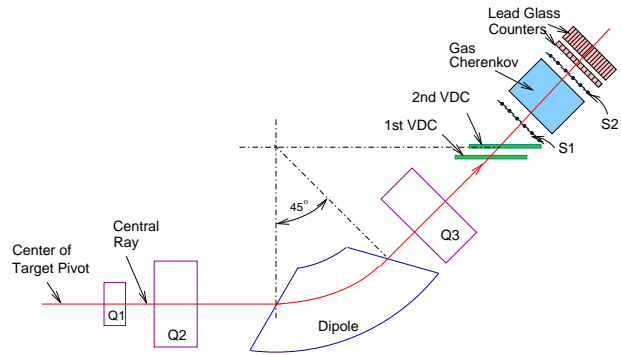}
 \end{center}
 \caption{(Color online) Schematic layout of the left HRS and detector package
(not to scale).}
\label{fig:hrs_side}
\end{figure}

The basic layout of the left HRS is shown in Fig.~\ref{fig:hrs_side}.  
The detector package is located in a large steel and concrete 
detector hut following the last magnet. For this experiment the detector 
package included (1) two scintillator planes S1 and S2 to provide a 
trigger to activate the DAQ electronics; (2) a set of two Vertical 
Drift Chambers (VDC)~\cite{exp:vdc} for particle tracking; 
(3) a gas $\breve{\q{C}}$erenkov detector
to provide particle identification (PID) information; and (4) a set of 
lead glass counters for additional PID.  The layout of the right HRS is 
almost identical except a slight difference in the geometry of the gas 
$\breve{\q{C}}$erenkov detector and the lead glass counters.\\

{\subsection{Particle Identification}}

For this experiment the largest background came from photo-produced
pions.  We refer to PID in this paper as the identification of electrons 
from pions. PID for each HRS was accomplished by a CO$_2$ threshold 
gas $\breve{\q{C}}$erenkov detector and a double-layered lead glass 
shower detector.  

The two $\breve{\q{C}}$erenkov detectors, one on each 
HRS, were operated  with CO$_2$ at atmospheric pressure. The refraction 
index of the CO$_2$ gas was 1.00041, giving a threshold momentum of 
$\approx 17$~MeV/c for electrons and $\approx 4.8$~GeV/c for pions. 
The incident particles on each HRS were also identified by their 
energy deposits in the lead glass shower detector.

\begin{figure}[htp]
 \begin{center}
  \includegraphics[scale=0.5]{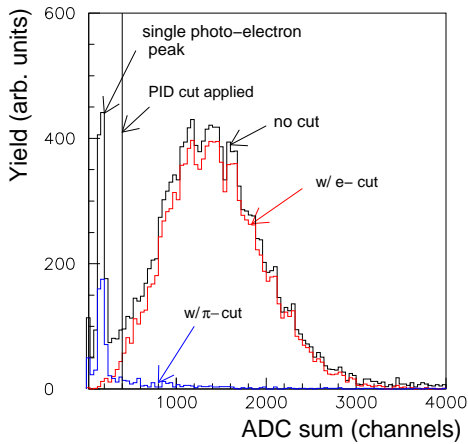}
 \end{center}
 \caption{(Color online) Summed ADC signal of the left HRS gas $\breve{\q{C}}$erenkov 
detector: without cuts, after lead glass counters electron 
cut and after pion cut. The vertical line shows a cut 
$\sum{\q{ADC}_i}>400$ applied to select 
 electrons.}\label{fig:leftgcadc}
\end{figure}

\begin{figure}[htp]
 \begin{center}
  \includegraphics[scale=0.6]{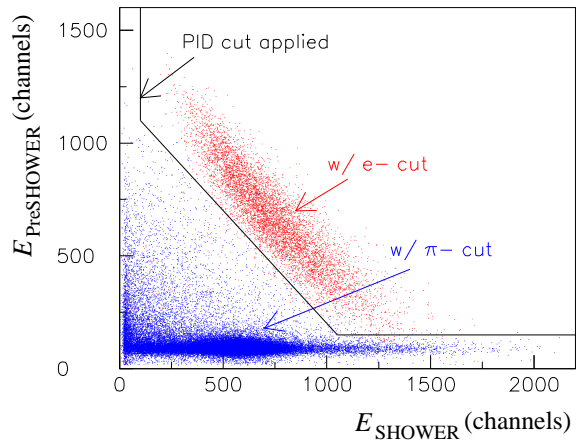}
 \end{center}
 \caption{(Color online) Energy deposited in the first layer (preshower) {\it vs}
that in the second layer (shower) of lead glass counters in the right HRS. 
The two blobs correspond to the spectrum with a tight gas 
$\breve{\q{C}}$erenkov ADC electron cut and with a pion cut applied. 
The lines show the boundary of the two-dimensional cut used to select 
electrons in the data analysis.}\label{fig:rightshadc}
\end{figure}

Since $\breve{\q{C}}$erenkov detectors and lead glass shower detectors 
are based on different mechanisms and their PID efficiencies are not 
correlated~\cite{book:partdet}, we extracted the PID efficiency of the 
lead glass counters by using electron events selected by the 
$\breve{\q{C}}$erenkov detector, and {\it vice versa}.
Fig.~\ref{fig:leftgcadc} shows a spectrum of the summed ADC signal of the 
left HRS gas $\breve{\q{C}}$erenkov detector, without a cut on the 
lead glass signal and after applying such lead glass electron and 
pion cuts. The spectrum from the right HRS is similar.
Fig.~\ref{fig:rightshadc} shows the distribution of the energy deposit 
in the two layers of the right HRS lead glass counters, without a
$\breve{\q{C}}$erenkov cut, and after $\breve{\q{C}}$erenkov electron 
and pion cuts.

Detailed PID analysis was done both before and during the experiment. 
The PID performance of each detector is characterized by the electron 
detection efficiency $\eta_{e}$ and the pion rejection factor 
$\eta_{\pi,\q{rej}}$, defined as the number of pions needed to cause 
one pion contamination event.  In the HRS central momentum range of 
$0.8<p_0<2.0$~(GeV/c), the PID efficiencies for the left HRS were 
found to be\\
%
 \indent$\diamond$ Gas $\breve{\q{C}}$erenkov:
 ~~{$\eta_{\pi,\q{rej}}>770$ at $\eta_{e}=99.9\%$;} \\
 \indent$\diamond$ Lead glass counters:
	{ $\eta_{\pi,\q{rej}} \approx 38$ at $\eta_{e}=98\%$;} \\
 \indent$\diamond$ Combined: $\eta_{\pi,\q{rej}} >3\times 10^{4}$ 
        at $\eta_{e}=98\%$.\\\\
%
and for the right HRS were\\
%
 \indent$\diamond$ Gas $\breve{\q{C}}$erenkov:
 ~~{$\eta_{\pi,\q{rej}}=900$ at $\eta_{e}=99\%$}; \\
  \indent$\diamond$ Lead glass counters: 
	{ $\eta_{\pi,\q{rej}} \approx 182$ at $\eta_{e}=98\%$;} \\
 \indent$\diamond$ Combined:
  $\eta_{\pi,\q{rej}} >1.6\times 10^{5}$ at $\eta_{e}=97\%$.\\

\subsection{Data Acquisition System}\label{ch3:daq}

We used the CEBAF Online Data Acquisition (CODA) system~\cite{exp:coda} 
for this experiment. In the raw data file, data from the detectors, 
the beamline equipment, and from the slow control software were recorded.
The total volume of data accumulated during the two-month running period 
was about 0.6~TBytes. 
Data from the detectors were processed using an analysis package 
called Experiment Scanning Program for hall A Collaboration Experiments 
(ESPACE)~\cite{exp:espace}. 
ESPACE was used to filter raw data, to make histograms for reconstructed
variables, to export variables into ntuples for further analysis, and to 
calibrate experiment-specific detector constants. It also provided the 
possibility to apply conditions on the incoming data. The information 
from scaler events was used to extract beam charge and DAQ deadtime 
corrections.

\section{The Polarized $\bm{^3\mathrm{He}}$ Target}\label{targ:main}

Polarized $^3$He targets are widely used at SLAC, DESY, MAINZ, 
MIT-Bates and JLab to study the electromagnetic structure and the 
spin structure of the neutron. There exist two major methods to 
polarize $^3$He nuclei.  The first one uses the metastable-exchange 
optical pumping technique~\cite{targ:metaexch}. The second method is
based on optical pumping~\cite{targ:optpump} and spin 
exchange~\cite{targ:spinexch}. It has been used at JLab since 
1998~\cite{exp:e94010}, and was used here.

The $^3\vec{\q{He}}$ target at JLab Hall~A uses the same design as 
the SLAC $^3\vec{\q{He}}$ target~\cite{thesis:romalis}. 
The first step to polarize $^3$He nuclei is to polarize an alkali 
metal vapor (rubidium was used at JLab as well as at SLAC) by optical 
pumping~\cite{targ:optpump} with circularly polarized laser light.
Depending on the photon helicity, the electrons in the Rb 
atoms will accumulate at either the $F=3,m_F=3$ or the $F=3,m_F=-3$ 
level (here $F$ is the atom's total spin and $m_F$ is its 
projection along the magnetic field axis). The polarization is then
transfered to the $^3$He nuclei through the spin exchange 
mechanism~\cite{targ:spinexch} during collisions between Rb atoms 
and the $^3$He nuclei.
Under operating conditions the $^3$He density is about $10^{20}$
nuclei/cm$^{3}$ and the Rb density is about $10^{14}$ atoms/cm$^{3}$.

To minimize depolarization effects caused by the unpolarized light 
emitted from decay of the excited electrons, N$_2$ buffer gas was 
added to provide a channel for the excited electrons to decay to 
the ground state without emitting photons~\cite{targ:optpump}. In 
the presence of N$_2$, electrons decay through collisions between the 
Rb atoms and N$_2$ molecules, which is usually referred to as non-radiative 
quenching. The number density of N$_2$ was about $1\%$ of that 
of $^3$He.

\subsection{Target Cells}\label{ch4:cell}
The target cells used for this experiment were 25-cm long 
pressurized glass cells with $\sim130$-$\mu$m thick end windows.
\begin{figure}[hbt]
 \begin{center}
  \includegraphics[width=230pt]{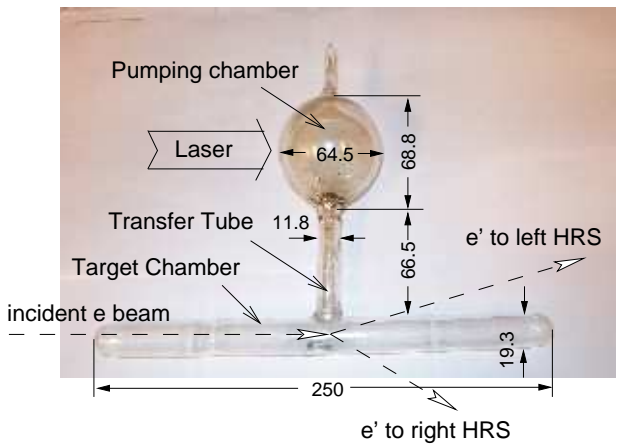}
 \end{center}
 \caption{(Color online) JLab target cell, geometries are given in mm for cell \#2
used in this experiment.} \label{tfig:targcell}
\end{figure}
\begin{table} [htp]
 \caption{Target cell characteristics. 
Symbols are: $V_p$ pumping chamber volume in cm$^3$;
 $V_t$ target chamber volume in cm$^3$;
 $V_{tr}$ transfer tube volume in cm$^3$;
 $V_0$ total volume in cm$^3$;
 $L_{tr}$ transfer tube length in cm;
 $n_0$: $^3$He density in amg at room temperature 
(1 amg $= 2.69\times 10^{-19}/$cm$^3$ which corresponds
to the gas density at the standard pressure and $T=0^\circ$C);
lifetime is in hours.
}
 \label{tab:cellchar}
  \begin{center}
 \begin{ruledtabular}
   \begin{tabular}{|c|cccc|c|c|c|}
 Name     & $V_p$ & $V_t$ & $V_{tr}$& $V_0$ &$L_{tr}$& $n_0$&lifetime\\\hline  
 Cell \#1 & 116.7 & 51.1  & 3.8     & 171.6 & 6.574  & 9.10 & 49 \\
 Cell \#2 & 116.1 & 53.5  & 3.9     & 173.5 & 6.46   & 8.28 & 44 \\
 uncertainty &1.5 & 1.0  & 0.25    & 1.8   & 0.020  & 2\%  & 1  \\
   \end{tabular}
\end{ruledtabular}
  \end{center}
\end{table}
The cell consisted of two chambers, a spherical upper chamber which 
holds the Rb vapor and in which the optical pumping occurs, and a long 
cylindrical chamber where the electron beam passes through and interacts 
with the polarized $^3$He nuclei. Two cells were used for this experiment. 
Figure~\ref{tfig:targcell} is a picture of the first cell with dimensions 
shown in mm. Table~\ref{tab:cellchar} gives the cell volumes and densities.

\subsection{Target Setup}\label{ch4:setup}
\begin{figure}[ht]
 \begin{center}
  \includegraphics[width=230pt]{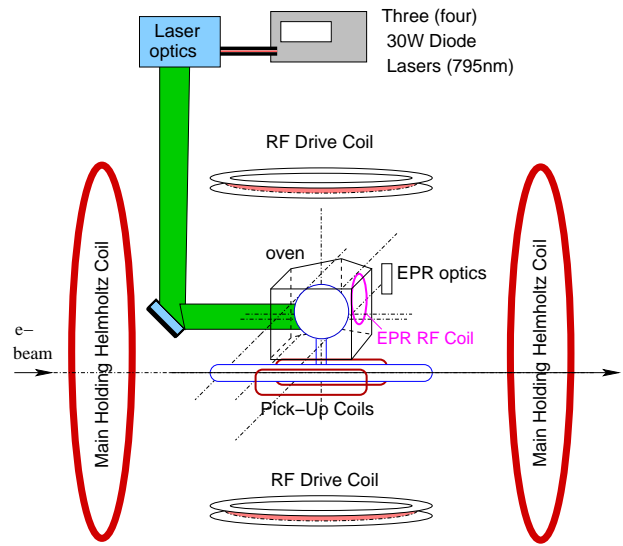}
 \end{center}
 \caption{(Color online) Target setup overview (schematic).}\label{tfig:overview}
\end{figure}

Figure~\ref{tfig:overview} is a schematic diagram of the target setup.  
There were two pairs of Helmholtz coils to provide a 25~G main 
holding field, with one pair oriented perpendicular and the other 
parallel to the beamline (only the perpendicular pair is shown).  The 
holding field could be aligned in any horizontal direction with respect to the 
incident electron beam.  The coils were excited by two power supplies 
in the constant voltage mode. The coil currents were continuously 
measured and recorded by the slow control system.
The cell was held at the center of the Helmholtz coils with its pumping chamber 
mounted inside an oven heated to $170^\circ$C in order to vaporize the Rb.
The lasers used to polarize the Rb were three 30~W diode lasers tuned to
a wavelength of 795~nm.  The target polarization was measured by two 
independent methods -- the NMR 
(Nuclear Magnetic Resonance)~\cite{exp:NIM,exp:e94010,thesis:kramer} 
and the EPR (Electro Paramagnetic 
Resonance)~\cite{exp:NIM,exp:e94010,PRA03004,thesis:zheng} polarimetry.
The NMR system consisted of one pair of pick-up coils (one on each side 
of the cell target chamber), one pair of RF coils and the associated 
electronics. The RF coils were placed at the top and the bottom of the 
scattering chamber, oriented in the horizontal plane, as shown in 
Fig.~\ref{tfig:overview}.  The EPR system 
shared the RF coils with the NMR system.  It consisted of one additional 
RF coil to induce light signal emission from the pumping chamber, a photodiode 
and the related optics to collect the light, 
and associated electronics for signal processing. 

\subsection{Laser System}\label{ch4:laser}
\begin{figure}[htp]
 \begin{center}
  \includegraphics[width=240pt, angle=0]{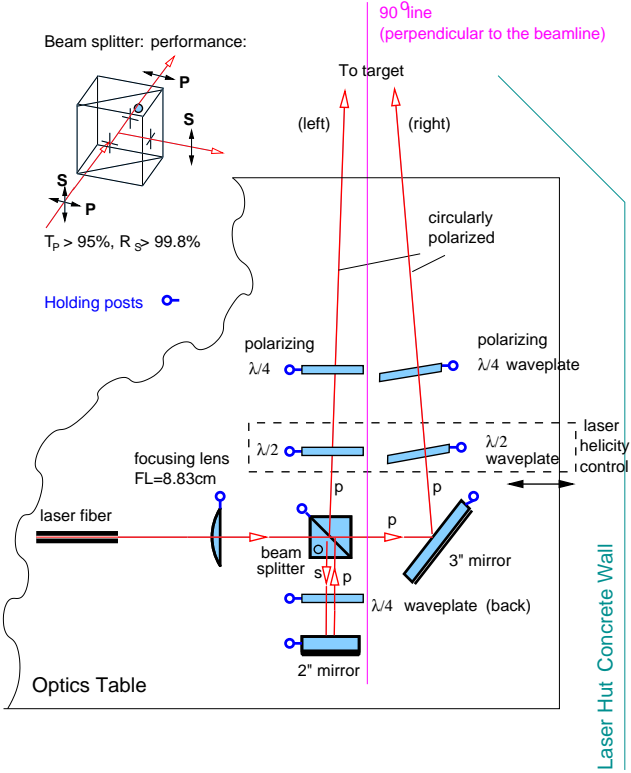}
 \end{center}
 \caption{(Color online) Laser polarizing optics setup (schematic) for the Hall~A 
polarized $^3$He target.}\label{tfig:lasersetup}
\end{figure}

The laser system used during this experiment consisted of seven diode 
lasers -- three for longitudinal pumping, three for transverse pumping 
and one spare. To protect the diode lasers from radiation damage from 
the electron beam, as well as to minimize the safety issues related to 
the laser hazard, the diode lasers and the associated optics system 
were located in a concrete laser hut located on the right side of the 
beamline at $90^\circ$, as shown in Fig.~\ref{fig:floorplan}.
The laser optics had seven individual lines, each associated with one 
diode laser.  All seven optical lines were identical and were placed 
one on top of the other on an optics table inside the laser hut.  
Each optical line consisted of one focusing lens to correct the angular
divergence of the laser beam, one beam-splitter to linearly polarize 
the lasers, two mirrors to direct them, three quarter waveplates to 
convert linear polarization to circular polarization, and two half
waveplates to reverse the laser helicity.  Figure~\ref{tfig:lasersetup} 
shows a schematic diagram 
of one optics line.  

Under the operating conditions for either longitudinal or transverse 
pumping, the original beam of each diode laser was divided into two by 
the beam-splitter. Therefore there were a total of six polarized laser 
beams entering the target. 
The diameter of each beam was about 5~cm which approximately matched 
the size of the pumping chamber. The target was about 5~m away from 
the optical table. For the pumping of the transversely polarized target, 
all these laser beams went 
directly towards the pumping chamber of the cell through a window on 
the side of the target scattering chamber enclosure.  For longitudinal 
pumping, they were guided towards the top of the scattering chamber, 
then were reflected twice and finally reached the cell pumping chamber.  

\subsection{NMR Polarimetry}\label{ch4:nmr}

The polarization of the $^3$He was determined by measuring the $^3$He 
Nuclear Magnetic Resonance (NMR) signal.  The principle of NMR polarimetry 
is the spin reversal of $^3$He nuclei using the Adiabatic Fast Passage 
(AFP)~\cite{targ:AFP} technique.  At resonance this spin reversal will 
induce an electromagnetic field and a signal in the pick-up coil pair. 
The signal magnitude is proportional to the polarization of the $^3$He and can be 
calibrated by performing the same measurement on a water sample, which 
measures the known thermal polarization of protons in water. The systematic 
error of the NMR measurement was about $3\%$, dominated by the error in 
the water calibration~\cite{thesis:kramer}.

\subsection{EPR Polarimetry}\label{ch4:epr}
In the presence of a magnetic field, the Zeeman splitting of Rb, 
characterized by the Electron-Paramagnetic Resonance frequency 
$\nu_\q{EPR}$, is proportional to the field magnitude.  When $^3$He 
nuclei are polarized ($P\approx 40\%$), their spins generate a small 
magnetic field $B_{^3\q{He}}$ of the order of $\approx 0.1$~Gauss, 
super-imposed on the main holding field $B_{H}=25$~Gauss. During an EPR 
measurement~\cite{PRA03004} the spin of the $^3$He is flipped by AFP, 
hence the direction of $B_{^3\q{He}}$ is reversed and the change 
in the total field magnitude causes a shift in $\nu_\q{EPR}$. This
frequency shift $\delta\nu_\q{EPR}$ is proportional to the $^3$He 
polarization in the pumping chamber.
The $^3$He polarization in the target chamber is calculated using
a model which describes the polarization diffusion from the pumping
chamber to target chamber.
The value of the EPR resonance frequency $\nu_\q{EPR}$ 
can also be used to calculate the magnetic field magnitude. The systematic 
error of the EPR measurement was about $3\%$, which came mainly from 
uncertainties in the cell density and temperature, and from the diffusion model.

\subsection{Target Performance}\label{ch4:polr}

\begin{figure}[htp]
 \begin{center}
  \includegraphics[angle=-90,scale=0.36]{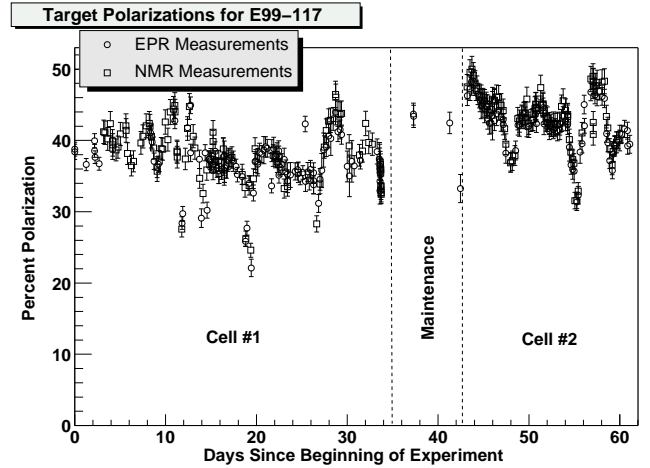}
 \end{center}
 \caption{Target polarization, starting June 1 of 2001, as measured by
EPR and NMR polarimetries.}\label{tfig:targperf}
\end{figure}

The target polarizations measured during this experiment are shown in 
Fig.~\ref{tfig:targperf}. Results from the two polarimetries are in 
good agreement and the average target polarization in beam 
was~$(40.0\pm 2.4)\%$. 
In a few cases the polarization measurement itself caused an abrupt
loss in the polarization. This phenomenon may be the so-called 
``masing effect''~\cite{thesis:romalis} due to non-linear couplings 
between the $^3$He spin rotation and conducting components inside 
the scattering chamber, {\it e.g.}, the NMR pick-up coils, and the 
``Rb-ring'' formed by the rubidium condensed inside the cell at 
the joint of the two chambers.  This masing effect was later 
suppressed by adding coils to produce an additional field gradient.

\section{Data Analysis}\label{ch5:main}

In this section we present the analysis procedure leading to the final 
results in Section~\ref{result:main}. We start with the analysis of
elastic scattering, the $\Delta(1232)$ transverse asymmetry, and the 
check for false asymmetry. Next, the DIS analysis and radiative 
corrections are presented.  Finally we describe nuclear corrections 
which were used to extract neutron structure functions from the $^3$He data.

\subsection{Analysis Procedure}\label{ch5:procedure}

\begin{figure}[htp]
  \begin{center}
  \includegraphics[scale=0.45]{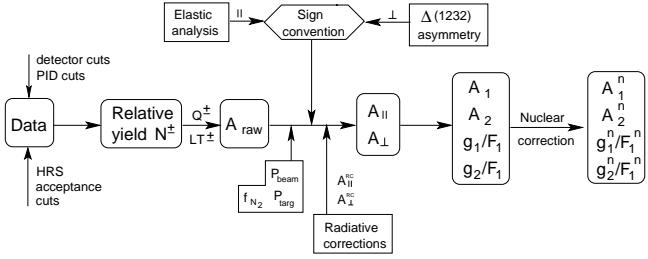}
  \end{center}
 \caption{Procedure for asymmetry analysis.}\label{fig:ana_procedure}
\end{figure}

The procedure to extract the electron asymmetries from our data is 
outlined in Fig.~\ref{fig:ana_procedure}.  
From the raw data one first obtains the helicity-dependent electron 
yield $N^{\pm}$ using acceptance and PID cuts.  The efficiencies 
associated with these cuts are not helicity-dependent, hence are 
not corrected for in the asymmetry analysis.  The yield is then 
corrected for the helicity-dependent integrated beam charge $Q^\pm$
and the livetime of the DAQ system $\q{LT}^\pm$.
The asymmetry of the corrected yield is the 
raw asymmetry $A_{raw}$. 
Next, to go from $A_{raw}$ to the 
physics asymmetries $A_{\parallel}$ and $A_{\perp}$, four factors 
need to be taken into account: the beam polarization $P_b$, the 
target polarization $P_t$, the nitrogen dilution factor $f_{\q{N}_2}$ 
due to the unpolarized nitrogen nuclei mixed with the polarized 
$^3$He gas, and a sign based on the knowledge of the absolute 
state of the electron helicity and the target spin direction: 
\begin{eqnarray}
 A_{\parallel,\perp} &=& \pm\frac{A_{raw}}{f_{\q{N}_2}P_bP_t}
  \label{equ:he3raw0} 
\end{eqnarray}
The results of the beam and the target polarization measurements 
have been presented in previous sections. The nitrogen dilution factor 
is obtained from data taken with
a reference cell filled with nitrogen.  The sign of the asymmetry
is described by ``the sign convention''.  The sign convention for 
parallel asymmetries was obtained from the elastic scattering 
asymmetry and that for perpendicular asymmetries was from 
the $\Delta(1232)$ asymmetry analysis, as will be described 
in Sections~\ref{ana:elana} and~\ref{ana:delta}.  
The physics asymmetries $A_{\parallel}$ and $A_{\perp}$, after 
corrections for radiative effects, were used to calculate 
$A_1$ and $A_2$ and the structure function ratios $g_1/F_1$ 
and $g_2/F_1$ using Eq.~(\ref{equ:a2a1}---\ref{equ:a2g2}). Then  
the last step is to apply nuclear corrections in order to extract 
the neutron asymmetries and the structure function ratios from the 
$^3$He results, as will be described in Section~\ref{ch5:he3model}.\\

\begin{figure}[htp]
  \begin{center}
  \includegraphics[scale=0.5]{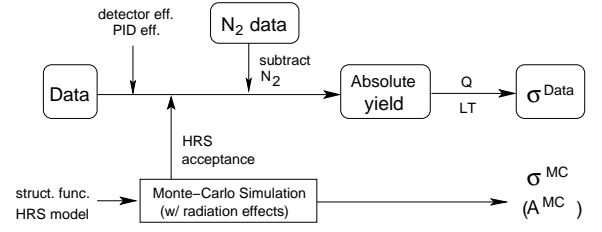}
  \end{center}
 \caption{Procedure for cross section analysis.}\label{fig:ana_xsec_procedure}
\end{figure}

Although the main goal of this experiment was to provide precise 
data on the asymmetries, cross sections were also extracted
from the data.  The procedure for the cross section analysis is outlined
in Fig.~\ref{fig:ana_xsec_procedure}.  One first determines the 
absolute yield of $\vec{e}-\vec{^3\q{He}}$ inclusive scattering 
from the raw data.  Unlike the asymmetry analysis, corrections need 
to be made for the detector and PID efficiencies and the spectrometer 
acceptance.  A Monte-Carlo simulation is used to calculate the spectrometer 
acceptance based on a transport model for the HRS~\cite{exp:NIM} with 
radiative effects taken into account. One then subtracts the yield of 
$e-\q{N}$ scattering caused by the N$_2$ nuclei in the target.  The clean 
$\vec{e}-\vec{^3\q{He}}$ yield is then corrected for the helicity-averaged 
beam charge and the DAQ livetime to give cross section results.  Using 
world fits for the unpolarized structure functions (form factors) 
of $^3$He, one can calculate the expected DIS (elastic) cross 
section from the Monte-Carlo simulation and compare to the data.

\subsection{Elastic Analysis}\label{ana:elana}
Data for $\vec{e}-^3\vec{\q{He}}$ elastic scattering were taken on 
a longitudinally polarized target with a beam energy of 1.2~GeV. The 
scattered electrons were detected at an angle of $20^\circ$.  The 
formalism for the cross sections and asymmetries are summarized in 
Appendix~\ref{app:el}.  Results for the elastic asymmetry 
were used to check the product of beam and target polarizations, as 
well as to determine the sign convention for different beam-helicity 
states and target spin directions.\\

The raw asymmetry was extracted from the data by
\begin{eqnarray}
 A_{raw} &=& \frac{\frac{N^+}{Q^+LT^+}-\frac{N^-}{Q^-LT^-}}
   {\frac{N^+}{Q^+LT^+}+\frac{N^-}{Q^-LT^-}} 
\label{equ:he3elaraw} 
\end{eqnarray}
with $N^{\pm}$, $Q^{\pm}$ and $LT^{\pm}$ the helicity-dependent 
yield, beam charge and livetime correction, respectively. 
The elastic asymmetry is
\begin{eqnarray}
 A_{\parallel}^{el} &=& \pm\frac{A_{raw}}{f_{\q{N}_2}f_{QE} P_bP_t}
  \label{equ:he3elaphys} 
\end{eqnarray}
with $f_{\q{N}_2}= 0.975\pm 0.003$ the N$_2$ dilution factor determined from
data taken with a reference cell filled with nitrogen,
and $P_b$ and $P_t$ the beam and target polarization, respectively.
A cut in the invariant mass $\vert{W-M_{^3\q{He}}}\vert < 6$~(MeV)
was used to select elastic events. Within this cut there are a small amount
of quasi-elastic events and $f_{QE}>0.99$ is the quasi-elastic dilution 
factor used to correct for this effect.

\begin{figure}[ht]
 \begin{center}
  \includegraphics[scale=0.4, angle=0]{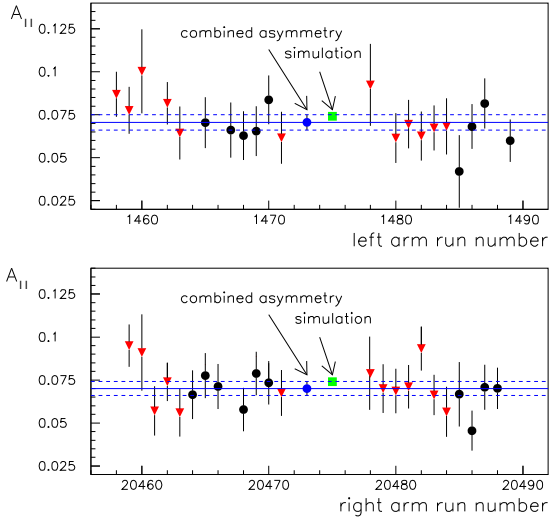}
 \end{center}
 \caption{(Color online) Elastic parallel asymmetry results for the two HRS. 
The kinematics are $E=1.2$~GeV and $\theta=20^\circ$. 
A cut in the invariant mass $\vert{W-M_{^3\q{He}}}\vert < 6$~(MeV)
was used to select elastic events. Data from runs with beam half-wave
plate inserted are shown as triangles. The error bars shown are total errors 
including a $4.5\%$ systematic uncertainty, which is dominated by the 
error of the beam and target polarizations. The combined asymmetry 
and its total error from $\approx 20$ runs are shown by the horizontal 
solid and dashed lines, respectively, as well as the solid circle 
as labeled~\cite{thesis:zheng}.}
\label{fig:elasym}
\end{figure}
The sign on the right hand side of Eq.~(\ref{equ:he3elaphys}) depends 
on the configuration of the beam half-wave plate, the spin precession
of electrons in the accelerator, and the target spin direction.
It was determined by comparing the sign of the measured raw asymmetries 
with the calculated elastic asymmetry.  We found that for this experiment
the electron helicity was aligned to the beam direction during H+ pulses
when the beam half-wave plate was {\it not} inserted. Since the electron 
spin precession in the accelerator can be well calculated using quantum
electro-dynamics and the results showed that the beam helicity during H+ 
pulses was the same for the two beam energies used for elastic and 
DIS measurements, the above convention also applies to the DIS data analysis.

\begin{figure}[ht]
 \begin{center}
  \includegraphics[scale=0.4, angle=0]{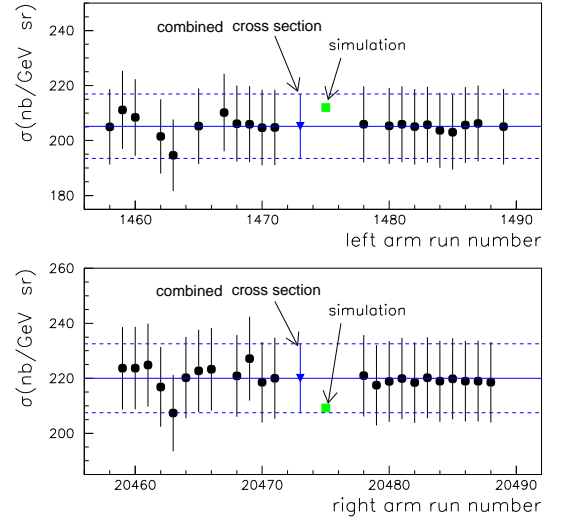}
 \end{center}
 \caption{(Color online) Elastic cross section results for the two HRS. The kinematics 
were $E=1.2$~GeV and $\theta=20^\circ$. A systematic error of $6.7\%$ 
was assigned to each data point, which was dominated by the uncertainty 
in the target density and the HRS transport functions~\cite{thesis:zheng}.}
\label{fig:elxsec}
\end{figure}

A Monte-Carlo simulation was performed which took into account the
spectrometer acceptance, the effect of the quasi-elastic scattering 
background and radiative effects. Results for the elastic 
asymmetry and the cross section are shown in Fig.~\ref{fig:elasym} 
and~\ref{fig:elxsec}, respectively, along with the expected values 
from the simulation. The data show good agreement with the simulation 
within the uncertainties.

\subsection{$\Delta (1232)$ Transverse Asymmetry}\label{ana:delta}

Data on the $\Delta(1232)$ resonance were taken on a transversely polarized
target using a beam energy of $1.2$~GeV. The scattered electrons were
detected at an angle of $20^\circ$ and the central momentum of the 
spectrometers was set to $0.8$~GeV/c.
The transverse asymmetry defined by Eq.~(\ref{equ:Aperpdef}) was extracted 
from the raw asymmetry using Eq.~(\ref{equ:he3raw0}).
\begin{figure}[ht]
 \begin{center}
  \includegraphics[scale=0.4, angle=0]{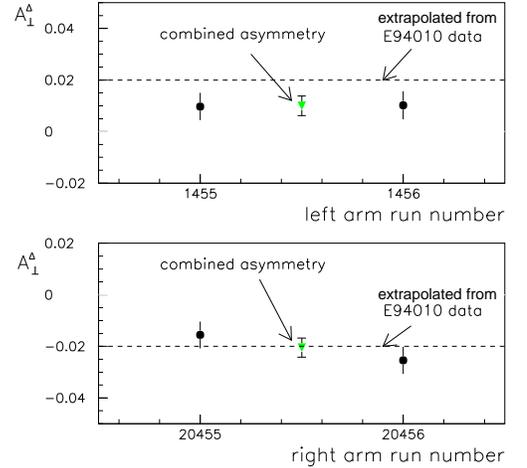}
 \end{center}
 \caption{(Color online) Measured raw $\Delta(1232)$ transverse asymmetry, with beam 
half-wave plate inserted and target spin pointing to the left side
of the beamline. The kinematics are $E=1.2$~GeV, $\theta=20^\circ$
and $E^\prime=0.8$~GeV/c. The dashed lines show the expected value 
obtained from previous $^3$He data extrapolated in $Q^2$.}
\label{fig:deltaasym}
\end{figure}
A cut in the invariant mass $\vert W-1232\vert < 20$~(MeV) 
was used to select $\Delta(1232)$ events. The sign on the right 
hand side of Eq.~(\ref{equ:he3raw0}) depends 
on the beam half-wave plate status, the spin precession
of electrons in the accelerator, the target spin direction, and in
which (left or right) HRS the asymmetry is measured.  Since data from
a previous experiment~\cite{exp:e94010} in a similar kinematic region
showed that $A^\Delta_\parallel<0$ and $A^\Delta_\perp>0$~\cite{thesis:deur}, 
$A^\Delta_\perp$ can be used to determine the sign convention of the 
measured transverse asymmetries.    
The raw $\Delta(1232)$ transverse asymmetry measured during this 
experiment was positive on the left HRS, as shown in Fig.~\ref{fig:deltaasym}, 
with the beam half-wave plate inserted and the target spin pointing to the left
side of the beamline. Also shown is the expected value obtained from
previous $^3$He data extrapolated in $Q^2$. Similar to the longitudinal 
configuration, this convention applied to both the $\Delta(1232)$ and DIS 
measurements.


\subsection{False Asymmetry and Background}\label{ana:false}

False asymmetries were checked by measuring the asymmetries 
from a polarized beam scattering off an unpolarized $^{12}$C 
target.  The results show that the false asymmetry was less than
$2\times 10^{-3}$, which
was negligible compared to the statistical uncertainties of
the measured $^3$He asymmetries.
To estimate the background from pair production $\gamma\to e^-+e^+$, 
the positron yield was measured at $x=0.33$, which is expected to have
the highest pair production background.  The positron cross section 
was found to be $\approx 3\%$ of the total cross section at $x=0.33$,
and the positron contribution at $x=0.48$ and $x=0.61$ should be even 
smaller. The effect of pair production asymmetry
is negligible compared to the statistical uncertainties of the measured 
$^3$He asymmetries and is not corrected for in this analysis.

\subsection{DIS Analysis}\label{ana:disana}

The longitudinal and transverse asymmetries defined by 
Eq.~(\ref{equ:Apardef}) and~(\ref{equ:Aperpdef}) for DIS were
extracted from the raw asymmetries as
\begin{eqnarray}
 A_{\parallel,\perp} &=&
  \pm\frac{A_{raw}}{f_{\q{N}_2} P_bP_t} \label{equ:aphysdis} 
\end{eqnarray}
where the sign on the right hand side was determined by the procedure 
described in Sections~\ref{ana:elana} and~\ref{ana:delta}.
The N$_2$ dilution factor, extracted from
runs where a reference cell was filled with pure N$_2$, was found
to be $f_{\q{N}_2} 0.938\pm 0.007$ for all three DIS kinematics.

\medskip
Radiative corrections were performed for the $^3$He asymmetries 
$A_\parallel^{^3\q{He}}$ and $A_\perp^{^3\q{He}}$. We denote by 
$A^{obs}$ the observed asymmetry, $A^{Born}$ the non-radiated 
(Born) asymmetry, $\Delta A^{ir}$ the correction due to internal 
radiation effects and $\Delta A^{er}$ the one due to external 
radiation effects. One has 
$A^{Born}=A^{obs}+\Delta A^{ir}+\Delta A^{er}$ 
for a specific target spin orientation. 

Internal corrections were calculated using an improved version of 
POLRAD~2.0~\cite{ana:polrad}. External corrections were calculated 
with a Monte-Carlo simulation based on the procedure first described 
by Mo and Tsai~\cite{ana:motsai}. Since the theory of radiative 
corrections is well established~\cite{ana:motsai}, the accuracy
of the radiative correction depends mainly on the structure functions 
used in the procedure. To estimate the uncertainty of both
corrections, five different 
fits~\cite{ana:f2comfst,ana:f2nmc92,ana:f2nmc95,ana:f2pslac94,ana:f2phallc02}
were used for the unpolarized structure function $F_2$ and two 
fits~\cite{ana:r1990,ana:r1998} were used for the ratio $R$.  For the 
polarized structure function $g_1$, in addition to those used in
POLRAD~2.0~\cite{ana:f2g1sch,ana:f2g1grsv96}, we fit to world $g_1^p/F_1^p$
and $g_1^n/F_1^n$ data including the new results from this experiment. 
Both fits will be presented in Section~\ref{result:neutron}. 
For $g_2$ we used both $g_2^{WW}$ and an assumption that $g_2=0$.
The variation in the radiative corrections using the fits 
listed above was taken as the full uncertainty of the corrections.
\begin{table}[ht]
 \caption{Total radiation length $X_0$ and thickness $d$ of the material 
traversed by incident (before interaction) and scattered (after 
interaction) electrons. The cell is made of glass
GE180 which has $X_0=7.04$~cm and density $\rho=2.77$~g/cm$^3$. 
The radiation length and thickness after interaction are
given by left/right depending on by which HRS the electrons were detected.}
\label{tab:radlen}
 \begin{center}
 \begin{ruledtabular}
 \begin{tabular}{c|c|c|c}
  $x$ &   0.33, 0.48 &   0.61 &  0.61 \\ \hline
  $\theta$ & 35$^\circ$ & 45$^\circ$ & 45$^\circ$ \\
  Cell &  \#2 & \#2 &  \#1    \\
  Cell window ($\mu$m) & 144 & 144 & 132 \\
  $X_0$ (before) &  0.00773 &  0.00773 & 0.00758 \\
  $d$ (g/cm$^2$, before)& 0.23479 & 0.23479 & 0.23317\\
  Cell wall (mm) &   1.44/1.33 & 1.44/1.33 & 1.34/1.43 \\
  $X_0$ (after) & 0.0444/0.0416 & 0.0376/0.0354 & 0.0356/0.0374\\
 $d$ (g/cm$^2$, after) & 0.9044/0.8506 & 0.7727/0.7293 & 0.7336/0.7687 \\ 
 \end{tabular}
 \end{ruledtabular}
 \end{center}
\end{table}
\begin{table}[!ht]
 \caption{Internal radiative corrections to $A_\parallel^{^3\q{He}}$ 
and $A_\perp^{^3\q{He}}$.
}
\label{tab:aIntRC}
 \begin{center}
 \begin{ruledtabular}
 \begin{tabular}{c|c|c}
  $x$ &
  $\Delta A_\parallel^{ir, ^3\q{He}}$ ($\times 10^{-3}$) &
  $\Delta A_\perp^{ir, ^3\q{He}}$ ($\times 10^{-3}$) \\ \hline
 0.33 & -5.77 $\pm$ 0.47 & 2.66 $\pm$ 0.03 \\ 
 0.48 & -3.28 $\pm$ 0.13 & 1.47 $\pm$ 0.05 \\ 
 0.61 & -2.66 $\pm$ 0.15 & 1.28 $\pm$ 0.07 \\ 
 \end{tabular}
 \end{ruledtabular}
 \end{center}
\end{table}
\begin{table}[!ht]
 \caption{External radiative corrections to $A_\parallel^{^3\q{He}}$ 
and $A_\perp^{^3\q{He}}$.
Errors are from uncertainties in the structure functions and in the 
cell wall thickness.
}
\label{tab:aExtRC}
 \begin{center}
 \begin{ruledtabular}
 \begin{tabular}{c|c|c}
  $x$ &
  $\Delta A_\parallel^{er,^3\q{He}}$ ($\times 10^{-3}$) &
  $\Delta A_\perp^{er,^3\q{He}}$ ($\times 10^{-3}$) \\ \hline
   0.33 & -0.67 $\pm$ 0.10 & -0.05 $\pm$ 0.11 \\ 
   0.48 & -1.16 $\pm$ 0.15 &  0.80 $\pm$ 0.46 \\ 
   0.61 & -0.39 $\pm$ 0.03 &  0.29 $\pm$ 0.04 \\ 
 \end{tabular}
 \end{ruledtabular}
 \end{center}
\end{table}
For external corrections the uncertainty also includes the contribution
from the uncertainty in the target cell wall thickness.
The total radiation length and thickness of the material traversed 
by the scattered electrons are given in Table~\ref{tab:radlen} for 
each kinematic setting.  Results for the internal and external radiative 
corrections are given in Table~\ref{tab:aIntRC} and~\ref{tab:aExtRC}, 
respectively.


By measuring DIS unpolarized cross sections and using the asymmetry
results, one can calculate the polarized cross sections and extract 
$g_1$ and $g_2$ from Eq.~(\ref{equ:polxsec_long}) and~(\ref{equ:polxsec_tran}).
We used a Monte-Carlo simulation to calculate the expected DIS 
unpolarized cross sections within the spectrometer acceptance. 
This simulation included internal and external radiative corrections. 
The structure functions 
used in the simulation were from the latest DIS world 
fits~\cite{ana:r1998,ana:f2nmc95} with the nuclear effects 
corrected~\cite{theory:wallyEMC}. The radiative corrections
from the elastic and quasi-elastic processes were calculated
in the peaking approximation~\cite{ana:peaking} using the 
world proton and neutron form factor 
data~\cite{ana:ffpro,ana:ffneu_dipole,ana:ffneu_galster}.  
The DIS cross section results agree with the simulation at a level 
of $10\%$.  Since this is not a dedicated cross section experiment,
we obtained the values for $g_1$ and $g_2$ by multiplying our
$g_1/F_1$ and $g_2/F_1$ results by the world fits for unpolarized 
structure functions $F_1$~\cite{ana:f2nmc95,ana:r1998}, instead of 
the $F_1$ from this analysis.

\subsection{From $^3$He to Neutron} \label{ch5:he3model}

Properties of protons and neutrons embedded in nuclei are 
expected to be different from those in free space because 
of a variety of nuclear effects, including that from 
spin depolarization, binding and Fermi motion, 
the off-shell nature of the nucleons, 
the presence of non-nucleonic degrees of freedom, and nuclear 
shadowing and antishadowing.
A coherent and complete picture of 
all these effects for the $^3$He structure function $g_1^{^3\q{He}}$ 
in the range of $10^{-4}\leq x\leqslant 0.8$ was presented 
in~\cite{theory:3Hecmplt}. It gives
\begin{eqnarray}\label{equ:he3-g1n}
 g_1^{^3\q{He}} &=& P_ng_1^n+2P_pg_1^p-0.014
   \big[{g_1^p(x)-4g_1^n(x)}\big]\nonumber \\
 && +a(x)g_1^n(x)+b(x)g_1^p(x)
\end{eqnarray}
where $P_n$($P_p$) is the effective polarization of the neutron 
(proton) inside $^3$He~\cite{theory:PnPp_friar}. Functions
$a(x)$ and $b(x)$ are $Q^2$-dependent and represent
the nuclear shadowing and antishadowing effects.

From Eq.(\ref{equ:a1=g1f1-simplified}), the asymmetry $A_1$ is 
approximately the ratio of the spin structure function $g_1$ and $F_1$.  
Noting that shadowing and antishadowing are not present in the large $x$ 
region, using Eq.~(\ref{equ:he3-g1n}) one obtains
\begin{eqnarray}\label{equ:he3ton}
 A_1^n&=&\frac{F_2^{^3\q{He}}
        \big[{A_1^{^3\q{He}}}-2\frac{F_2^p}{F_2^{^3\q{He}}}
	  P_pA_1^p(1-\frac{0.014}{2P_p})\big]}
        {P_nF_2^n(1+\frac{0.056}{P_n})}~. 
\end{eqnarray}
The two terms $\frac{0.056}{P_n}$ and $\frac{0.014}{2P_p}$ represent 
the corrections to $A_1^n$ associated with the $\Delta(1232)$ component 
in the $^3$He wavefunction. Both terms cause $A_1^n$ to increase
in the $x$ range of this experiment, and to turn 
positive at lower values of $x$ compared to the situation when the 
effect of the $\Delta(1232)$ is ignored. For $F_2^n$ and $F_2^{^3\q{He}}$, 
we used the world proton and deuteron $F_2$ data and took into account
the EMC effects~\cite{theory:wallyEMC}. We used the world proton asymmetry 
data for $A_1^p$.  The effective nucleon polarizations $P_{n,p}$ can be
calculated using $^3$He wavefunctions constructed from N-N interactions,
and their uncertainties were estimated using various nuclear 
models~\cite{theory:PnPp_nogga,theory:PnPp_friar,theory:3Heconv,theory:PnPp_bissey},
giving 
\begin{eqnarray}
&& P_n=0.86^{+0.036}_{-0.02}~~\q{and}~P_p=-0.028^{+0.009}_{-0.004}~.\label{equ:PnPp}
\end{eqnarray}
Eq.~(\ref{equ:he3ton}) was also used for extracting 
$A_2^n$, $g_1^n/F_1^n$ and $g_2^n/F_1^n$ from our $^3$He data.
The uncertainty in $A_1^n$ due to the uncertainties in $F_2^{p,d}$,
in the correction for EMC effects, in $A_1^p$ data and 
in $P_{n,p}$ is given in Table~\ref{tab:errana}.
Compared to the convolution approach~\cite{theory:3Heconv}
used by previous $^3$He 
experiments~\cite{data:a1ng1n-e142,data:a1ng1n-e154,data:a1ng1n-hermes}, 
in which only the first two terms on the right hand side of 
Eq.~(\ref{equ:he3-g1n}) are present, the values of $A_1^n$
extracted from Eq.~(\ref{equ:he3ton}) are larger by $(1-2)\%$ in the 
region $0.2<x<0.7$. \\

\medskip
\subsection{Resonance Contributions}
Since there are a few nucleon resonances with masses above 2~GeV and
our measurement at the highest $x$ point has an invariant mass close
to $2$~GeV, the effect of possible contributions from baryon resonances 
were evaluated. This was done by comparing the resonance contribution
to $g_1^n$ with that to $F_1^n$.  For our kinematics at $x=0.6$, data 
on the unpolarized structure function $F_2$ and $R$~\cite{data:f2p-hallc} 
show that the resonance contribution to $F_1$ is less than $5\%$.
The resonance asymmetry was estimated using the MAID model~\cite{theory:maid} and was found 
to be approximately $0.10$ at $W=1.7$~(GeV). Since the resonance structure is 
more evident at smaller $W$, we took this value as an upper limit of the 
contribution at $W=2$~(GeV).
The resonance contribution to our $A_1^n$ and $g_1^n/F_1^n$ results at 
$x=0.6$ were then estimated to be at most $0.008$,
which is negligible compared to their statistical errors.


%
\section{{RESULTS}}\label{result:main}

\subsection{$^3$He Results}\label{result:he3}

Results of the electron asymmetries for $\vec{e}-^3\vec{\q{He}}$ scattering, 
$A_\parallel^{^3\q{He}}$ and $A_\perp^{^3\q{He}}$, 
the virtual photon asymmetries $A_{1}^{^3\q{He}}$ 
and $A_{2}^{^3\q{He}}$, structure function ratios 
$g_{1}^{^3\q{He}}/F_1^{^3\q{He}}$ and $g_{2}^{^3\q{He}}/F_1^{^3\q{He}}$
and polarized structure functions $g_{1}^{^3\q{He}}$
and $g_{2}^{^3\q{He}}$ are given in Table~\ref{tab:result_he3}.  
Results for $g_{1,2}^{^3\q{He}}$ were obtained by multiplying the
$g_{1,2}^{^3\q{He}}/F_1^{^3\q{He}}$ results by the unpolarized 
structure function $F_1^{^3\q{He}}$, which were calculated using 
the latest world fits of DIS data~\cite{ana:r1998,ana:f2nmc95} 
and with nuclear effects corrected~\cite{theory:wallyEMC}. 
Results for $A_1^{^3\q{He}}$ and $g_1^{^3\q{He}}$ are shown 
in Fig.~\ref{fig:result_a1he3} along with 
SLAC~\cite{data:a1ng1n-e142,data:a1heg1he-e154} 
and HERMES~\cite{data:hermes_dqq} data.\\

\begin{widetext}

\begin{figure*}[!ht]
\vspace*{0.4cm}
  \includegraphics[angle = 0, width=237pt]{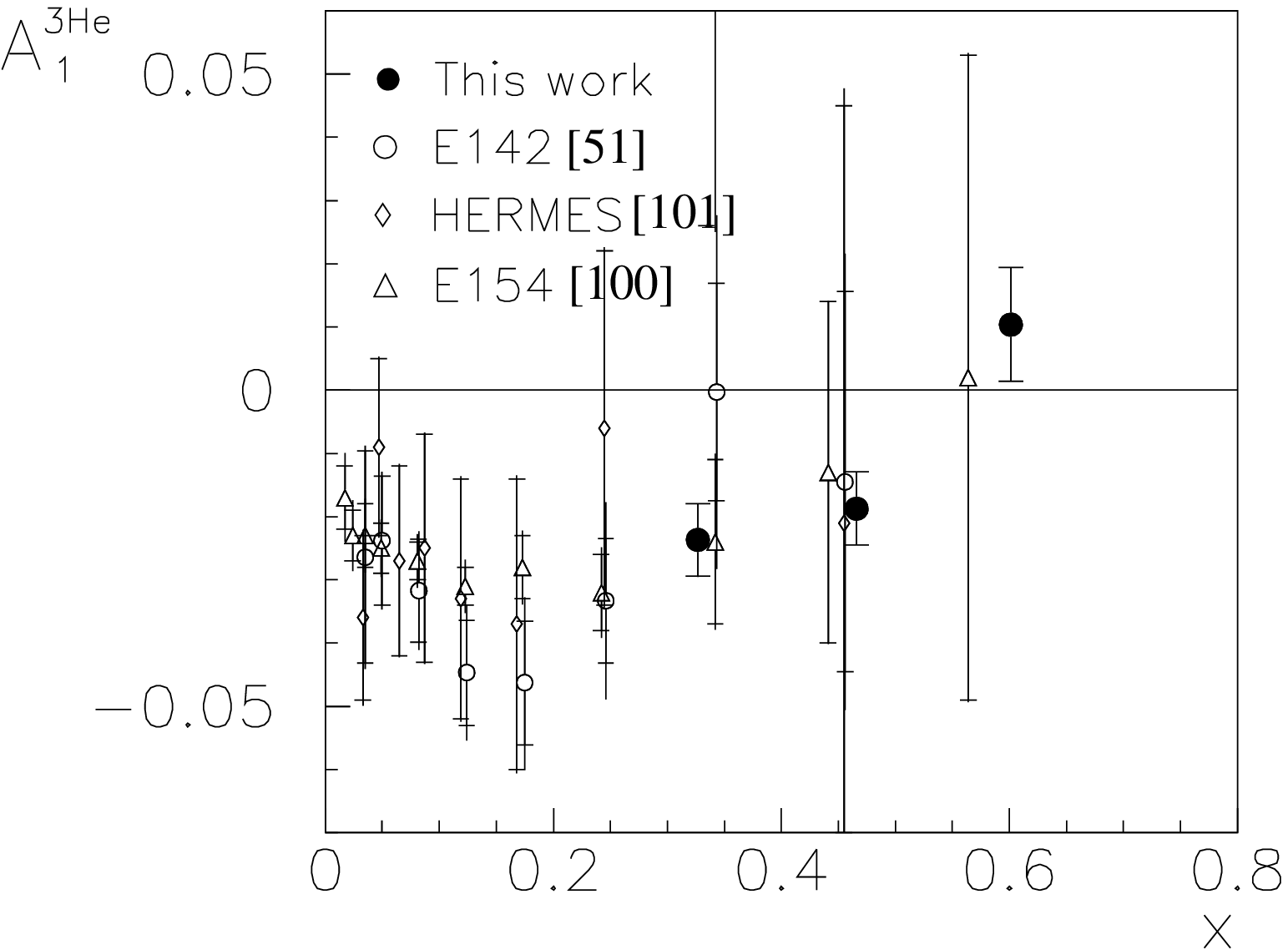}
 \hspace*{0.6cm}
  \includegraphics[angle = 0, width=225pt]{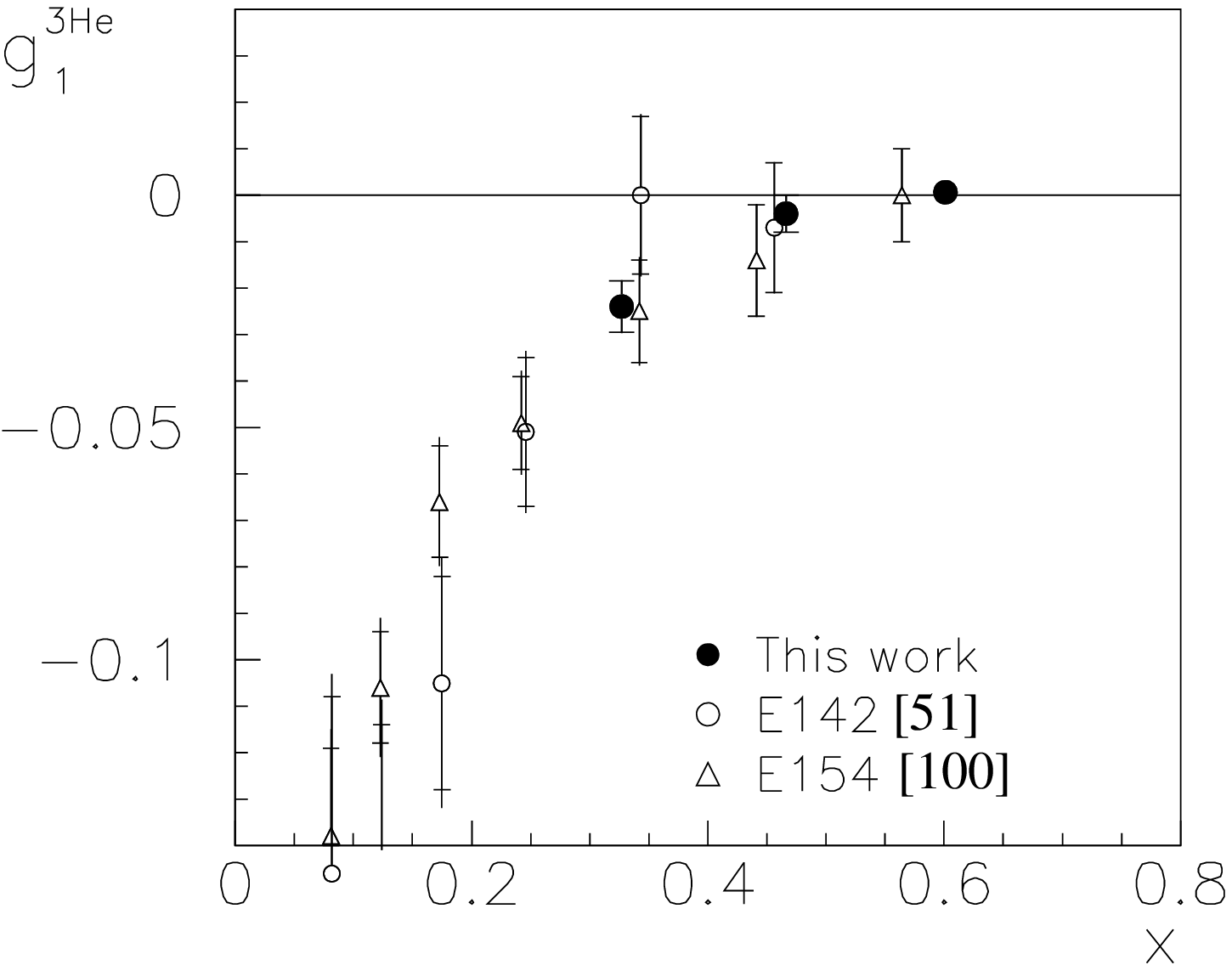}
 \caption{Results for the $^3$He asymmetry $A_1^{^3\q{He}}$ 
and the structure functions $g_1^{^3\q{He}}$ as a function of $x$, along
with previous data from SLAC~\cite{data:a1ng1n-e142,data:a1heg1he-e154} 
and HERMES~\cite{data:hermes_dqq}. Error bars of the results from this work
include both statistical and systematic uncertainties.}\label{fig:result_a1he3}
\end{figure*}
\begin{table*}[!ht]
\caption{\label{tab:result_he3} Results for $^3$He 
asymmetries $A_1^{^3\q{He}}$ and $A_2^{^3\q{He}}$,
structure function ratios $g_1^{^3\q{He}}/F_1^{^3\q{He}}$ 
and $g_2^{^3\q{He}}/F_1^{^3\q{He}}$, and polarized structure 
functions $g_1^{^3\q{He}}$ and $g_2^{^3\q{He}}$.
Errors are given as $\pm$ statistical $\pm$ systematic.
}
\begin{ruledtabular}
\begin{tabular}{c|r|r|r}
 $\langle x \rangle$              & 0.33 & 0.47 & 0.60 \\
 $\langle Q^2\rangle$ (GeV/c)$^2$ & 2.71 & 3.52 & 4.83 \\ \hline
 $A_\parallel^{^3\q{He}}$ 
    & $-0.020\pm 0.005\pm 0.001$ 
    & $-0.012\pm 0.005\pm 0.000$
    & $ 0.007\pm 0.007\pm 0.001$  \\ 
 $A_\perp^{^3\q{He}}$
    & $ 0.000\pm 0.010\pm 0.000$
    & $ 0.016\pm 0.008\pm 0.001$
    & $-0.010\pm 0.016\pm 0.001$ \\
 $A_1^{^3\q{He}}$
    & $-0.024\pm 0.006\pm 0.001$ 
    & $-0.019\pm 0.006\pm 0.001$ 
    & $ 0.010\pm 0.009\pm 0.001$ \\
 $A_2^{^3\q{He}}$
    & $-0.004\pm 0.014\pm 0.001$ 
    & $ 0.020\pm 0.012\pm 0.001$  
    & $-0.013\pm 0.023\pm 0.001$ \\
 $g_1^{^3\q{He}}/F_1^{^3\q{He}}$
    & $-0.022\pm 0.005\pm 0.001$
    & $-0.008\pm 0.008\pm 0.001$
    & $ 0.003\pm 0.009\pm 0.001$ \\
 $g_2^{^3\q{He}}/F_1^{^3\q{He}}$
    & $ 0.010\pm 0.036\pm 0.002$ 
    & $ 0.050\pm 0.022\pm 0.003$ 
    & $-0.028\pm 0.038\pm 0.002$ \\
 $g_1^{^3\q{He}}$
    & $-0.024\pm 0.006\pm 0.001 $ 
    & $-0.004\pm 0.004\pm 0.000 $ 
    & $ 0.001\pm 0.002\pm 0.000 $  \\ 
 $g_2^{^3\q{He}}$
    & $ 0.011\pm 0.039\pm 0.001 $
    & $ 0.026\pm 0.012\pm 0.002 $
    & $-0.006\pm 0.009\pm 0.001 $ \\ 
\end{tabular}
\end{ruledtabular}
\end{table*}
%

\vspace*{1cm}
\subsection{Neutron Results}\label{result:neutron}
Results for the neutron asymmetries $A_1^n$ and $A_2^n$, structure 
function ratios $g_1^n/F_1^n$ and $g_2^n/F_1^n$ and polarized structure 
functions $g_1^n$ and $g_2^n$ are given in Table~\ref{tab:result_neutron}. 

\renewcommand{\arraystretch}{1.4}
\begin{table*}[!ht]
\begin{center}
\caption{Results for the asymmetries and spin structure functions
for the neutron. Errors are given as $\pm$statistical $\pm$systematic.
}
\label{tab:result_neutron}
\begin{ruledtabular}
\begin{tabular}{c|r|r|r}
 $\langle x \rangle$               & 0.33 & 0.47 & 0.60 \\
 $\langle Q^2\rangle$ (GeV/c)$^2$ & 2.71 & 3.52 & 4.83 \\ \hline
 $A_1^{n}$
    & $ -0.048\pm  0.024^{+ 0.015}_{-0.016}$
    & $ -0.006\pm  0.027^{+ 0.019}_{-0.019}$
    & $  0.175\pm  0.048^{+ 0.026}_{-0.028}$ \\ 
 $A_2^{n}$
    & $ -0.004\pm  0.063^{+ 0.005}_{-0.005}$
    & $  0.117\pm  0.055^{+ 0.012}_{-0.021}$
    & $ -0.034\pm  0.124^{+ 0.014}_{-0.014}$ \\ 
 $g_1^n/F_1^n$
    & $ -0.043\pm  0.022^{+ 0.009}_{-0.009}$
    & $  0.040\pm  0.035^{+ 0.011}_{-0.011}$
    & $  0.124\pm  0.045^{+ 0.016}_{-0.017}$ \\ 
 $g_2^n/F_1^n$
    & $  0.034\pm  0.153^{+ 0.010}_{-0.010}$
    & $  0.207\pm  0.103^{+ 0.022}_{-0.021}$
    & $ -0.190\pm  0.204^{+ 0.027}_{-0.027}$ \\ 
 $g_1^{n}$
    & $ -0.012\pm  0.006^{+ 0.003}_{-0.003}$
    & $  0.005\pm  0.004^{+ 0.001}_{-0.001}$
    & $  0.006\pm  0.002^{+ 0.001}_{-0.001}$ \\ 
 $g_2^{n}$
    & $  0.009\pm  0.043^{+ 0.003}_{-0.003}$
    & $  0.026\pm  0.013^{+ 0.003}_{-0.003}$
    & $ -0.009\pm  0.009^{+ 0.001}_{-0.001}$ \\
\end{tabular}
\end{ruledtabular}
\end{center}
\end{table*}
\renewcommand{\arraystretch}{1.0}
\end{widetext}

\begin{figure}[ht]
 \begin{center}
 \includegraphics[angle = 0, width=225pt]{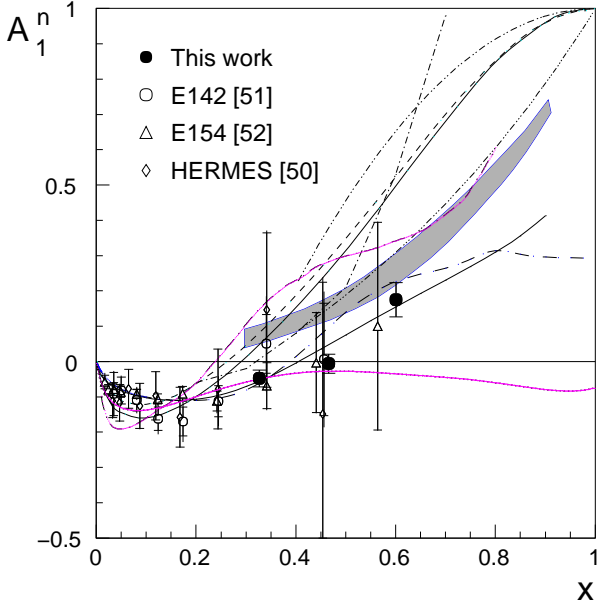}
 \end{center}
\caption{Our $A_1^n$ results along with theoretical predictions 
and previous world data obtained from polarized $^3$He 
targets~\cite{data:a1ng1n-e142,data:a1ng1n-e154,data:a1ng1n-hermes}. 
Curves: 
predictions of $A_1^n$ from SU(6) symmetry ($x$ axis at zero)~\cite{theory:su6close},
constituent quark model (shaded band)~\cite{theory:cqm},
statistical model at $Q^2=4$~(GeV/c)$^2$ (long-dashed)~\cite{theory:stat},
quark-hadron duality using two different SU(6) breaking mechanisms 
(dash-dot-dotted and dash-dot-dot-dotted), 
and non-meson cloudy bag model (dash-dotted)~\cite{theory:bag};
predictions of ${g_1^n}/{F_1^n}$ from pQCD HHC based BBS 
parameterization at $Q^2=4$~(GeV/c)$^2$ (higher solid)~\cite{theory:bbs} and LSS(BBS) 
parameterization at $Q^2=4$~(GeV/c)$^2$ (dashed)~\cite{theory:lssbbs}, LSS~2001 NLO polarized parton 
densities at $Q^2=5$~(GeV/c)$^2$ (lower solid)~\cite{theory:lss2001} and 
chiral soliton 
models~\cite{theory:chi_weigel} at $Q^2=3$~(GeV/c)$^2$ (long dash-dotted)
and~\cite{theory:chi_waka} at $Q^2=4.8$~(GeV/c)$^2$ (dotted).
}
\label{fig:a1nresult_all}
\end{figure}

\begin{figure}[!htp]
\vspace*{0.4cm}
  \includegraphics[angle = 0, width=225pt]{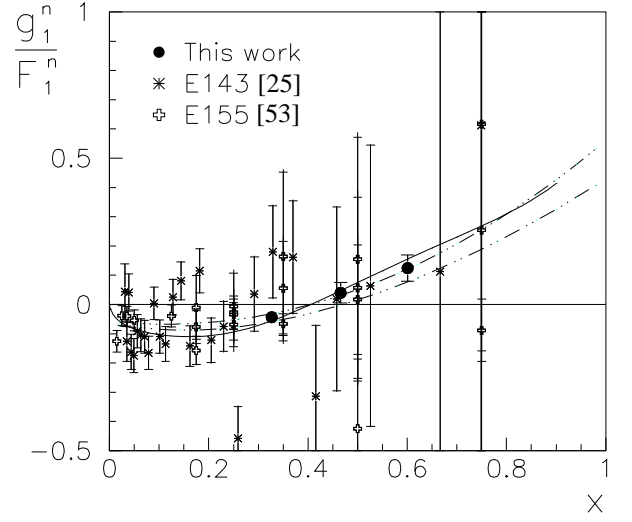}
 \caption{Results for $g_1^n/F_1^n$ along with previous world
data from SLAC~\cite{data:a1pa1n-e143,data:g1pg1n-e155}. The curves
are the prediction for $g_1^n/F_1^n$ from the LSS~2001 NLO polarized 
parton densities at $Q^2=5$~(GeV/c)$^2$~\cite{theory:lss2001},
the E155 experimental fit at $Q^2=5$~(GeV/c)$^2$ 
(long dash-dot-dotted)~\cite{data:g1pg1n-e155} 
and the new fit as described in the text (long dash-dot-dot-dotted).
}
\label{fig:result_a1n1}
\end{figure}

\begin{figure}[!htp]
\vspace*{0.4cm}
  \includegraphics[angle = 0, width=225pt]{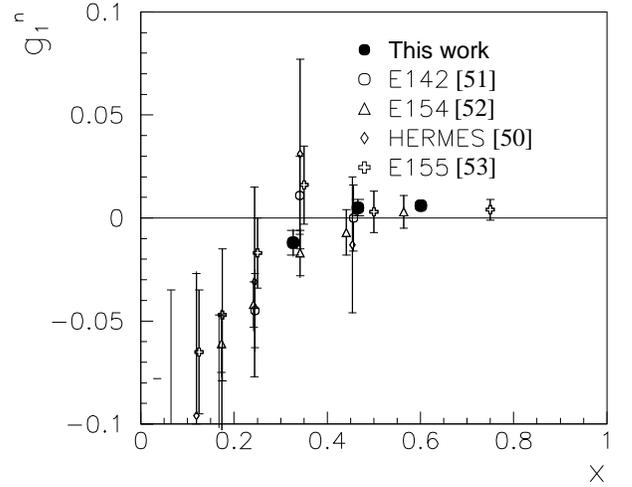}
 \caption{Results for $g_1^n$ along with previous world data from 
SLAC~\cite{data:a1ng1n-e142,data:a1ng1n-e154,data:g1pg1n-e155} 
and HERMES~\cite{data:a1ng1n-hermes}.}\label{fig:result_g1n}
\end{figure}

The $A_1^n$, $g_1^n/F_1^n$ and $g_1^n$ results are shown in 
Fig.~\ref{fig:a1nresult_all}, \ref{fig:result_a1n1} and \ref{fig:result_g1n},
respectively.  In the region of $x>0.4$, our results have improved
the world data precision by about an order of magnitude, and will provide
valuable inputs to parton distribution function (PDF) parameterizations.
Our data at $x=0.33$ are in good agreement with previous
world data. For the $A_1^n$ results, this is the first time that the 
data show a clear trend that $A_1^n$ turns to positive values at large $x$. 
As $x$ increases, the agreement between the data and the predictions from
the constituent quark models (CQM) becomes better.
This is within the expectation since the CQM is more likely to work in the 
valence quark region. It also indicates that $A_1^n$ will go to higher 
values at $x>0.6$.  However, the trend of the $A_1^n$ results does not agree 
with the BBS and LSS(BBS) parameterizations, which are from leading-order 
pQCD analyses based on hadron helicity conservation (HHC). This 
indicates that there might be problem in the assumption that quarks have
zero orbital angular momentum which is used by HHC.

%
%


The sources for the experimental systematic uncertainties are 
listed in Table~\ref{tab:expsysErr}.
\begin{table}[!ht]
\begin{center}
\caption{Experimental systematic errors for the $A_1^n$ result.}
\label{tab:expsysErr}
\begin{ruledtabular}
\begin{tabular}{l|r}
 source                           &  error      \\ \hline
Beam energy          $E_{b}$      & ${\Delta E_b}/{E_b}<5\times 10^{-4}$ \\
HRS central momentum $p_0$        & ${\Delta E_e}/{E_e}<5\times 10^{-4}$~\cite{technote:HRSgamma} \\
HRS central angle    $\theta_0$   & $\Delta \theta_0<0.1^\circ$~\cite{exp:hallATN02-032} \\
Beam polarization    $P_b$        & ${\Delta P_b}/{P_b}<3\%$ \\
Target polarization  $P_t$        & ${\Delta P_t}/{P_t}<4\%$ \\
Target spin direction $\alpha_t$  & $\Delta\alpha_t <1^\circ$ \\ 
\end{tabular}
\end{ruledtabular}
\end{center}
\end{table}
Systematic uncertainties for the $A_1^n$ results include that 
from experimental systematic errors, uncertainties in internal radiative 
corrections $\Delta A_1^{n,ir}$ and external radiative corrections
$\Delta A_1^{n,er}$ as derived from the values in
Tables~\ref{tab:aIntRC} and~\ref{tab:aExtRC}, and that from nuclear corrections
as described in Section~\ref{ch5:he3model}.
Table~\ref{tab:errana} gives these systematic uncertainties for the $A_1^n$
results along with their statistical uncertainties.
The total uncertainties are dominated by the statistical uncertainties.\\
\renewcommand{\arraystretch}{1.}
\begin{table}[!ht]
\begin{center}
\caption{Total uncertainties for $A_1^n$.
}
\label{tab:errana}
\begin{ruledtabular}
\begin{tabular}{l|r|r|r}
 $\langle x \rangle$ & 0.33   & 0.47   & 0.60   \\ \hline
Statistics           & 0.024 & 0.027 & 0.048 \\ 
Experimental syst.   & 0.004 & 0.003 & 0.004 \\ 
$\Delta A_1^{n,ir}$  & 0.012 & 0.013 & 0.015 \\ 
$\Delta A_1^{n,er}$  & 0.002 & 0.002 & 0.003  \\ 
$F_2^p$, $F_2^d$     & 0.006 & 0.008 & $^{+0.005}_{-0.010}$ \\ 
EMC effect           & 0.001 & 0.000 & 0.009 \\ 
$A_1^p$              & 0.001 & 0.005 & 0.011 \\ 
$P_n$, $P_p$         & $^{+0.005}_{-0.012}$ & $^{+0.009}_{-0.020}$ & $^{+0.018}_{-0.037}$ \\ 
\end{tabular}
\end{ruledtabular}
\end{center}
\end{table}
\renewcommand{\arraystretch}{1.0}
%

We used five functional forms, $x^\alpha P_n(x) (1+\beta/Q^2)$,
to fit our $g_1^n/F_1^n$ results combined 
with data from previous experiments~\cite{data:a1pa1n-e143,data:g1pg1n-e155}.
Here $P_n$ is the $n^{th}$-order polynomial, $n=1,2$ for a finite $\alpha$ or
$n=1,2,3$ if $\alpha$ is fixed to be $0$. The total number of parameters is limited
to $\leqslant 5$. For the $Q^2$-dependence of $g_1/F_1$, we used a term $1+\beta/Q^2$
as in the E155 experimental fit~\cite{data:g1pg1n-e155}.
No constraints were imposed
on the fit concerning the behavior of $g_1/F_1$ as $x\to 1$.
The function which gives the smallest
$\chi^2$ value is $g_1^n/F_1^n = (a+bx+cx^2) (1+\beta/Q^2)$. The new fit
is shown in Fig.~\ref{fig:result_a1n1}. Results for
the fit parameters are given in Table~\ref{tab:g1f1nfitpar}
and the covariance error matrix is
\begin{displaymath}
 {\epsilon} =
  \left [ \begin{array}{cccc}
  1.000 &  -0.737 &   0.148 &   0.960 \\
 -0.737 &   1.000 &  -0.752 &  -0.581 \\ 
  0.148 &  -0.752 &   1.000 &  -0.039 \\ 
  0.960 &  -0.581 &  -0.039 &   1.000 
  \end{array} \right ]~.
\end{displaymath}
\begin{table}[h]
 \caption{Result of the fit $g_1^n/F_1^n = (a+bx+cx^2) (1+\beta/Q^2)$.}
 \label{tab:g1f1nfitpar}
 \begin{center}
 \begin{tabular}{l} \hline
      \hspace{0.5cm} $a = -0.049 \pm 0.052$  \hspace{1cm}\\ 
      \hspace{0.5cm} $b = -0.162 \pm 0.217$ \\ 
      \hspace{0.5cm} $c = 0.698 \pm 0.345$ \\ 
      \hspace{0.5cm} $\beta = 0.751 \pm 2.174$ \\ \hline
 \end{tabular}
 \end{center}
\end{table}

Similar fits were performed to the proton world 
data~\cite{data:g1p-hermes,data:a1pa1n-e143,data:g1pg1n-e155} and
function $g_1^p/F_1^p = x^\alpha (a+bx) (1+\beta/Q^2)$ was found to
give the smallest $\chi^2$ value. The new fit is shown in 
Fig.~\ref{fig:a1pmodel_all} of Section~\ref{ch2:exp}. Results for 
the fit parameters are given in Table~\ref{tab:g1f1pfitpar} 
and the covariance error matrix is\\
\begin{displaymath}
 {\epsilon} =
  \left [ \begin{array}{cccc}
  1.000 &   0.908 &  -0.851 &   0.723 \\
  0.908 &   1.000 &  -0.967 &   0.401 \\
 -0.851 &  -0.967 &   1.000 &  -0.369 \\
  0.723 &   0.401 &  -0.369 &   1.000 
  \end{array} \right ]~. 
\end{displaymath} 
\begin{table}[h]
 \caption{Result of the fit $g_1^p/F_1^p = x^\alpha (a+bx) (1+\beta/Q^2)$.}
 \label{tab:g1f1pfitpar}
 \begin{center}
 \begin{tabular}{l} \hline
   \hspace{0.5cm}   $\alpha = 0.813 \pm 0.049$  \hspace{1cm}\\
     \hspace{0.5cm} $a = 1.231 \pm 0.122$ \\
     \hspace{0.5cm} $b = -0.413 \pm 0.216$ \\ 
     \hspace{0.5cm} $\beta = 0.030 \pm 0.124$ \\ \hline 
 \end{tabular}
 \end{center}
\end{table}
%

\begin{figure}[!htp]
\vspace*{0.4cm}
 \includegraphics[angle = 0, width=225pt]{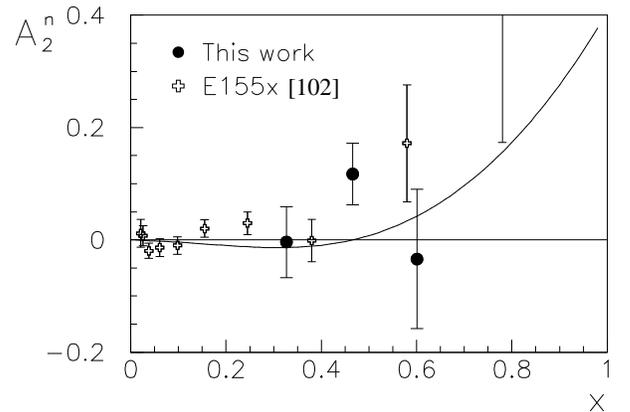}
 \caption{Results for $A_2^n$ along with the best previous 
world data~\cite{data:e155x}. The curve gives the twist-2 contribution
at $Q^2=4$~(GeV/c)$^2$ calculated using the E155 experimental 
fit~\cite{data:g1pg1n-e155} and $g_2^{WW}$ of Eq.~(\ref{equ:g2ww}).}
\label{fig:result_a2n}
\end{figure}

\begin{figure}[!htp]
\vspace*{0.4cm}
 \includegraphics[angle = 0, width=235pt]{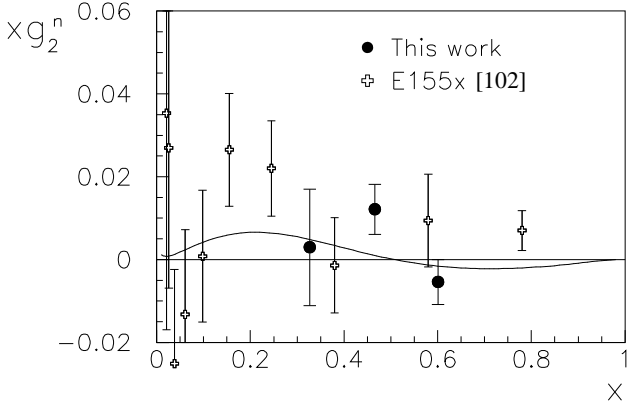}
 \caption{Results for $xg_2^n$ along with the best previous world 
data~\cite{data:e155x}. The curve gives the twist-2 contribution
at $Q^2=4$~(GeV/c)$^2$ calculated using the E155 experimental 
fit~\cite{data:g1pg1n-e155} and $g_2^{WW}$ of Eq.~(\ref{equ:g2ww}).}
\label{fig:result_xg2n}
\end{figure}

Figures~\ref{fig:result_a2n} and~\ref{fig:result_xg2n} show the results
for $A_2^n$ and $xg_2^n$, respectively. The precision of our data is comparable 
to the data from E155x experiment at SLAC~\cite{data:e155x}, which is so far 
the only experiment dedicated to measuring $g_2$ with published results. 

\smallskip
To evaluate the matrix element $d_2^n$, we combined our $g_2^n$ results 
with the E155x data~\cite{data:e155x}. 
The average $Q^2$ of the E155x data set is about $5$~(GeV/c)$^2$.
Following a similar procedure as used in
Ref.~\cite{data:e155x}, we assumed that $\bar g_2(x,Q^2)$ is independent 
of $Q^2$ and $\bar g_2\propto(1-x)^m$ with $m=2$ or $3$ for $x\gtorder 0.78$ 
beyond the measured region of both experiments. We obtained from Eq.~(\ref{equ:d2def})
\begin{eqnarray}
 d_2^n &=& 0.0062\pm 0.0028~.
\end{eqnarray}
Compared to the value published previously~\cite{data:e155x}, the 
uncertainty on $d_2^n$ has been improved by about a factor of two.
The large decrease in uncertainty despite the small number of our
data points arises from the $x^2$ weighting of the integral which
emphasizes the large $x$ kinematics. The 
uncertainties on the integrand has been improved in the region $x>0.4$ 
due to our $g_2^n$ results at the two higher $x$ points being more 
precise than that of E155x.
While a negative value was predicted by lattice QCD~\cite{theory:d2lattice}
and most other models~\cite{theory:d2bag,theory:d2QCDSR,theory:d2chi}, 
the new result for $d_2^n$ suggests that the higher twist contribution is positive.

%

\subsection{Flavor Decomposition using the Quark-Parton Model}\label{result:dqq}

Assuming the strange quark distributions $s(x)$, $\bar{s}(x)$, 
$\Delta s(x)$ and $\Delta \bar{s}(x)$ to be small in the 
region $x>0.3$, and ignoring any $Q^2$-dependence of the ratio of
structure functions, one can 
extract polarized quark distribution functions based on the 
quark-parton model as  
\begin{eqnarray}
 {{\frac{\Delta u+\Delta\bar u}{u+\bar u}}} &=&
	\frac{4 g_1^p(4+\duratio)}{15 F_1^p}
	-\frac{g_1^n(1+4\duratio)}{15 F_1^n}~
 \label{equ:duu}
\end{eqnarray}
and
\begin{eqnarray}
 {{\frac{\Delta d+\Delta\bar d}{d+\bar d}}} &=&
	\frac{4 g_1^n(1+4\duratio)}{15 F_1^n\duratio}
	-\frac{g_1^p(4+\duratio)}{15 F_1^p\duratio} ~,
\label{equ:ddd}
\end{eqnarray}
with $\duratio\equiv ({d+\bar d})/({u+\bar u})$.
Results for $({\Delta u+\Delta\bar u})/(u+$ $\bar u)$ and 
$({\Delta d+\Delta\bar d})/({d+\bar d})$ are given in Table~\ref{tab:result_dqq}.
As inputs we used our own results for $g_1^n/F_1^n$, the world data on 
$g_1^p/F_1^p$~\cite{thesis:zheng}, and the ratio $\duratio$ extracted from 
proton and deuteron unpolarized structure function data~\cite{theory:duratio}.
In a similar manner as for Eq.(\ref{equ:duu}) and (\ref{equ:ddd}) and ignoring 
nuclear effects, one can also add the world data on 
$g_1^{^2\q{H}}/F_1^{^2\q{H}}$ to the fitted data set and
extract these polarized quark distributions.
The results are, however, consistent with 
those given in Table~\ref{tab:result_dqq} and 
have very similar error bars
because the data 
on the deuteron in general have poorer precision than the data on the proton
and the neutron data from this experiment. 
The results presented here have changed compared to the values published previously
in Ref.~\cite{A1nPRL} due to an error discovered 
in our fitting of $\duratio$ from Ref.~\cite{theory:duratio}.
The analysis procedure is consistent with what was used in Ref.~\cite{A1nPRL}.%
\renewcommand{\arraystretch}{1.25}
\begin{table}[h]
\begin{center}
\caption{Results for the polarized quark distributions.
The three uncertainties are those due to the $g_1^n/F_1^n$ statistical 
error, $g_1^n/F_1^n$ systematic uncertainty and the uncertainties of 
the $g_1^p/F_1^p$ data, the $\duratio$ fit and the correction for 
$s$ and $c$ quark contributions.}
\label{tab:result_dqq}
\begin{ruledtabular}
\begin{tabular}{c|r|r}
 $\langle x \rangle$  & $({\Delta u+\Delta\bar u})/({u+\bar u})$ 
  & $({\Delta d+\Delta\bar d})/({d+\bar d})$  \\ \hline
 $0.33$ & $ 0.545 \pm 0.004 \pm 0.002 _{-0.025 }^{+0.024}$ 
        & $-0.352 \pm 0.035 \pm 0.014 _{-0.031 }^{+0.017}$\\
 $0.47$ & $ 0.649 \pm 0.006 \pm 0.002 _{-0.058 }^{+0.058}$ 
        & $-0.393 \pm 0.063 \pm 0.020 _{-0.049 }^{+0.041}$\\
 $0.60$ & $ 0.728 \pm 0.006 \pm 0.002 _{-0.114 }^{+0.114}$ 
        & $-0.440 \pm 0.092 \pm 0.035 _{-0.142 }^{+0.107}$\\
\end{tabular}
\end{ruledtabular}
\end{center}
\end{table}
\renewcommand{\arraystretch}{1.0}
%
\begin{figure}[h!]
 \includegraphics[angle = 0, width=225pt]{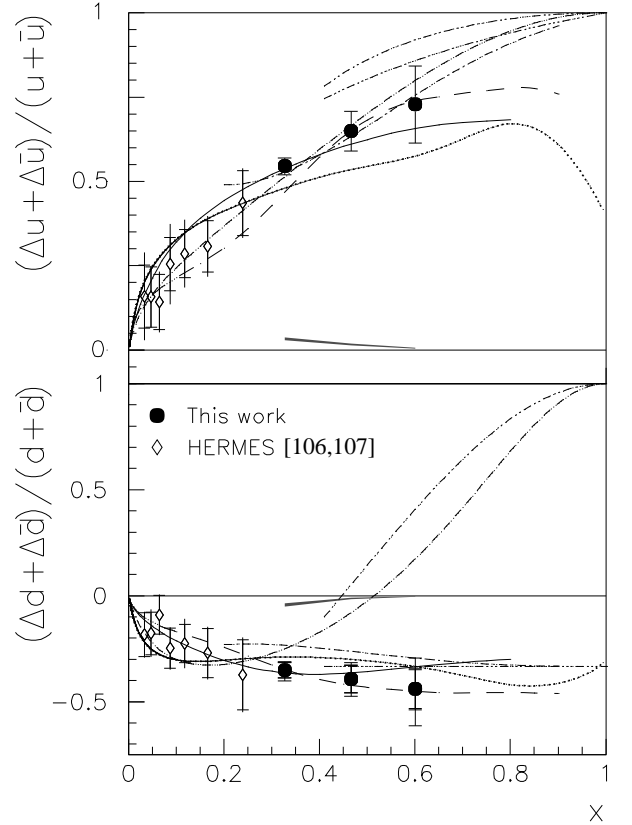}
 \caption{Results for $({\Delta u+\Delta\bar u})/({u+\bar u})$ and 
$({\Delta d+\Delta\bar d})/({d+\bar d})$ in the quark-parton model, 
compared with semi-inclusive data from HERMES~\cite{data:hermes_newdqq}
and CTEQ unpolarized PDF~\cite{theory:cteq} as described in the text, the RCQM 
predictions (dash-dotted)~\cite{theory:cqm}, predictions from LSS~2001 NLO 
polarized parton densities at $Q^2=5$~(GeV/c)$^2$ (solid)~\cite{theory:lss2001}, 
the statistical model at $Q^2=4$~(GeV/c)$^2$ (long-dashed)~\cite{theory:stat}, 
the pQCD-based predictions with the HHC constraint (dashed)~\cite{theory:lssbbs}, 
the duality model using two different SU(6) breaking mechanisms 
(dash-dot-dotted and dash-dot-dot-dotted)~\cite{theory:dual_new}, 
and predictions from chiral soliton model at $Q^2=4.8$~(GeV/c)$^2$ 
(dotted)~\cite{theory:chi_waka}.
The error bars of our data include the uncertainties given in 
Table~\ref{tab:result_dqq}.  
The shaded band near the horizontal axis shows the difference between 
$\Delta q_V/q_V$ and $(\Delta q+\Delta\bar q)/(q+\bar q)$ that needs to 
be added to the data when comparing with the RCQM calculation.}
\label{fig:result_dqq}
\end{figure}
%

Figure~\ref{fig:result_dqq} shows our results along with semi-inclusive data
on $(\Delta q+\Delta\bar q)/(q+\bar q)$ obtained from recent results for 
$\Delta q$ and $\Delta \bar q$~\cite{data:hermes_newdqq} by the HERMES 
collaboration, and the CTEQ6M unpolarized PDF~\cite{theory:cteq}.
To estimate the effect of the $s$ and $\bar s$ contributions,
we used two unpolarized PDF sets, CTEQ6M~\cite{theory:cteq} and 
MRST2001~\cite{theory:mrst}, and three polarized PDF sets, 
AAC2003~\cite{theory:aac03_polpdf}, BB2002~\cite{theory:bb_polpdf}
and GRSV2000~\cite{theory:grsv2000_polpdf}. For $c$ and $\bar c$ contributions
we used the two unpolarized PDF sets~\cite{theory:cteq,theory:mrst} and the 
positivity conditions that $\vert{\Delta c/c}\vert\leqslant~1$ and
$\vert{\Delta\bar c/\bar c}\vert\leqslant~1$.
To compare with the RCQM predictions, which are given for valence quarks, 
the difference between $\Delta q_V/q_V$ and $(\Delta q+\Delta\bar q)/(q+\bar q)$
was estimated using the two unpolarized PDF sets~\cite{theory:cteq,theory:mrst} 
and the three polarized PDF
sets~\cite{theory:aac03_polpdf,theory:bb_polpdf,theory:grsv2000_polpdf}
and is shown as the shaded band near the horizontal axis of 
Fig.~\ref{fig:result_dqq}. Here $q_V$($\Delta q_V$) is the unpolarized (polarized) 
valence quark distribution for $u$ or $d$ quark.  
Results shown in Fig.~\ref{fig:result_dqq} agree well with the 
predictions from the RCQM~\cite{theory:cqm} and the LSS 2001 NLO polarized 
parton densities~\cite{theory:lss2001}.  The results agree reasonably 
well with the statistical model calculation~\cite{theory:stat}.  
But results for the $d$ quark do not agree with the predictions 
from the leading-order pQCD LSS(BBS) parameterization~\cite{theory:lssbbs}
assuming hadron helicity conservation.

\section{{ CONCLUSIONS}}

We have presented precise data on the neutron spin asymmetry $A_1^n$ 
and the structure function ratio $g_1^n/F_1^n$ in the deep inelastic 
region at large $x$ obtained from a polarized $^3$He target.
These results will provide valuable inputs to the QCD parameterizations
of parton densities.
The new data show a clear trend that $A_1^n$ becomes positive at large~$x$.
Our results for $A_1^n$ agree with the LSS 2001 NLO QCD fit to the previous 
data and the trend of the $x$-dependence of $A_1^n$ agrees with the 
hyperfine-perturbed RCQM predictions. 
Data on the transverse asymmetry and structure function $A_2^n$ and
$g_2^n$ were also obtained with a precision comparable to the best previous 
world data in this kinematic region. 
Combined with previous world data,
the matrix element $d_2^n$ was evaluated and the new value differs from zero
by more than two standard deviations. This result suggests that the higher
twist contribution is positive.
Combined with the world proton data, the polarized quark distributions 
$(\Delta u+\Delta \bar u)/(u+\bar u)$ and 
$(\Delta d+\Delta\bar d)/(d+\bar d)$ were extracted based on the quark
parton model.  While results for $(\Delta u+\Delta \bar u)/(u+\bar u)$ 
agree well with predictions from various models and fits to the previous 
data, results for $(\Delta d+\Delta\bar d)/(d+\bar d)$ agree
with the predictions from RCQM and from the LSS 2001 fit, but do 
not agree with leading order pQCD predictions that use 
hadron helicity conservation. 
Since hadron helicity conservation is based on the assumption that
quarks have negligible orbital angular momentum, 
the new results suggest that the quark 
orbital angular momentum, or other effects beyond leading-order pQCD, 
may play an important role in this kinematic region.
\\


\appendix

\section{Formalism for Electron Deep Inelastic Scattering}\label{app:formalism}

\begin{figure}[htp]
 \begin{center}
 \includegraphics[angle=0,width=120pt]{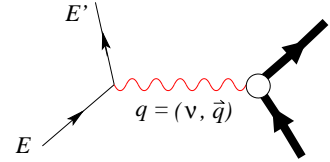} 
 \caption{(Color online) Electron scattering in the one-photon exchange approximation.}
 \label{fig:eNscat}
 \end{center}
\end{figure}

The fundamental quark and gluon structure of strongly 
interacting matter is studied primarily through experiments 
that emphasize hard scattering from the quarks and gluons 
at sufficiently high energies.  One important way of probing 
the distribution of quarks and antiquarks inside the nucleon is
electron scattering, where an electron scatters from a 
single quark or antiquark inside the target nucleon 
and transfers a large fraction of its energy and 
momentum via exchanged photons.  In the single photon exchange 
approximation, the electron interacts 
with the target nucleon via only one photon, as shown in 
Fig.~\ref{fig:eNscat}~\cite{book:thomas&weise}, and probes the
quark structure of the nucleon with a spatial resolution 
determined by the four momentum 
transfer squared of the photon $Q^2\equiv -q^2$.
Moreover, if a polarized electron beam and a polarized target
are used, the spin structure of the nucleon becomes accessible.
In the following we denote the incident electron energy by $E$, the 
energy of the scattered electron by $E^\prime$ thus the energy transfer 
of the photon is $\nu=E-E^\prime$, and the three-momentum 
transfer from the electron to the target nucleus by $\vec{q}$. 

\subsection{Structure Functions}\label{ch1:unpolstrf}

In the case of unpolarized electrons scattering off an 
unpolarized target, the differential cross-section for 
detecting the outgoing electron in a solid angle $\q{d}\Omega$ 
and an energy range ($E^\prime$, $E^\prime+\q{d}E^\prime$) 
in the laboratory frame can be written as
\begin{eqnarray} \label{equ:dis_xsec_f1f2}
 \frac{\mathrm{d}^2\sigma}{\mathrm{d}\Omega\mathrm{d}E^\prime} 
    \hspace*{-0.0cm}= &&\hspace*{-0.3cm}
    \Big(\frac{\mathrm{d}\sigma}{\mathrm{d}\Omega}\Big)_{Mott} \cdot
    \nonumber\\
    &&\Big[\frac{1}{\nu}F_2(x,Q^2)
    +\frac{2}{M}F_1(x,Q^2)\tan^2\frac{\theta}{2}\Big]~,
\end{eqnarray}
where $\theta$ is the scattering angle of the electron in the 
laboratory frame. The four momentum transfer $Q^2$ is given by
\begin{eqnarray}
 Q^2&=&4EE^\prime\sin^2\frac{\theta}{2} \label{equ:qmu2}~,
\end{eqnarray}
and the Mott cross section,
\begin{eqnarray}
{\Big(\frac{\mathrm{d}\sigma}{\mathrm{d}\Omega}\Big)}_{Mott}
 &=&\frac{\alpha^2\cos^2{\frac{\theta}{2}}}
 {4E^2\sin^4{\frac{\theta}{2}}}
 = \frac{\alpha^2\cos^2{\frac{\theta}{2}}}
 {Q^4}\frac{E^\prime}{E}~ \label{equ:Mottxsec}
\end{eqnarray}
with $\alpha$ the fine structure constant,
is the cross section for scattering 
relativistic electrons from a spin-0 point-like infinitely heavy target.
$F_1(x,Q^2)$ and $F_2(x,Q^2)$ are the unpolarized structure 
functions of the target, which are related to each other as
\begin{eqnarray}
 F_1(x,Q^2) &=& \frac{F_2(x,Q^2)(1+\gamma^2)}
                {2x\Big(1+R(x,Q^2)\Big)} \label{equ:disF1F2}
\end{eqnarray}
with $\gamma^2={(2Mx)^2}/{Q^2}$.
Here $R$ is defined as $R\equiv {\sigma_L}/{\sigma_T}$
with $\sigma_L$ and $\sigma_T$ the 
longitudinal and transverse virtual photon cross sections,
which can also be expressed in terms of $F_1$ and $F_2$.

Note that for a nuclear target, there exists an alternative 
{\it per nucleon} definition~({\it e.g.} as used in Ref.~\cite{ana:f2nmc95})
which is $1/A$ times the definition 
used in this paper, here $A$ is the number of nucleons inside 
the target nucleus.\\

A review of doubly polarized DIS was given in Ref.~\cite{theory:disreview}.
When the incident electrons are longitudinally polarized, the cross section 
difference between scattering off a target with its nuclear (or nucleon) spins 
aligned anti-parallel and parallel to the incident electron momentum is
\begin{eqnarray}
 &&\hspace*{-0.5cm} \frac{\q{d}^2\sigma_{\ua\Da}}{\q{d}\Omega
 \q{d}E^\prime}-\frac{\q{d}^2\sigma_{\ua\Ua}}{\q{d}\Omega \q{d}E^\prime}
 = \frac{4\alpha^2 E^\prime}{\nu EQ^2} \nonumber\\
 &&\times\Big [{(E+E^\prime\cos{\theta}){{g_1(x,Q^2)}}
   -2Mx{{g_2(x,Q^2)}}}\Big ]~. \label{equ:polxsec_long}
\end{eqnarray}
where $g_1 (x,Q^2)$ and $g_2(x,Q^2)$ are the polarized structure functions.
If the target nucleons are transversely polarized, then the cross section
difference is given by
\begin{eqnarray}
 \frac{\q{d}^2\sigma_{\ua\Ra}}{\q{d}\Omega \q{d}E^\prime}
  -&&\hspace*{-0.4cm}\frac{\q{d}^2\sigma_{\ua\La}}
  {\q{d}\Omega \q{d}E^\prime} 
 = \frac{4\alpha^2 E^{^\prime2}}{\nu EQ^2}\sin{\theta}\nonumber\\
 &&\hspace*{-0.cm}\times\Big [{{g_1(x,Q^2)}}+\frac{2E}{\nu}
   {{g_2(x,Q^2)}}\Big ]~. \label{equ:polxsec_tran}
\end{eqnarray}

\subsection{Bjorken Scaling and Its Violation}\label{ch1:bjscaling}

A remarkable feature of the structure functions $F_1$, $F_2$,
$g_1$ and $g_2$ is their scaling behavior. 
In the Bjorken limit~\cite{theory:bjorken}
($Q^2\to \infty$ and $\nu\to\infty$ at a fixed value of $x$), 
the structure functions become independent 
of $Q^2$~\cite{data:slac-bjscaling}. Moreover, in this limit 
$\sigma_L$ vanishes~\cite{book:thomas&weise}, hence $R=0$ 
and Eq.~(\ref{equ:disF1F2}) reduces to $F_2(x)=2xF_1(x)$,
known as the Callan-Gross relation~\cite{theory:Callan-Gross}.

At finite $Q^2$, the scaling of structure functions is violated
due to the radiation of gluons by both initial and scattered quarks. 
These gluon radiative corrections cause a logarithmic 
$Q^2$-dependence to the structure functions, which has been
verified by experimental data~\cite{exp:pdg} and can be precisely 
calculated in pQCD using the 
Dokshitzer-Gribov-Lipatov-Altarelli-Parisi (DGLAP)
evolution equations~\cite{theory:dglap}.

\subsection{From Bjorken Limit to Finite $Q^2$ using
the Operator Product Expansion}\label{ch1:ope}

In order to calculate observables at finite values of $Q^2$, a method called 
the Operator Product Expansion (OPE)~\cite{theory:ope} can be applied 
to DIS which can separate the non-perturbative part of an observable 
from its perturbative part.  In the OPE, whether an operator is 
perturbative or not is characterized by the ``twist'' of the operator.  
At large $Q^2$ the leading twist $t=2$ term dominates, while at small 
$Q^2$ higher-twist operators need to be taken into account, which are 
sensitive to interactions beyond the quark-parton model, {\it e.g.}, 
quark-gluon and quark-quark correlations~\cite{theory:ope-g2}.

\subsection{Virtual Photon-Nucleon Asymmetries}\label{ch1:a1ndef}

Virtual photon asymmetries are defined in terms of a 
helicity decomposition of the virtual photon absorption
cross sections~\cite{theory:drechsel}. For 
the absorption of circularly polarized virtual photons 
with helicity $\pm 1$ by longitudinally polarized nucleons,
the longitudinal asymmetry $A_1$ is defined as
\begin{eqnarray}
 A_1(x,Q^2) &\equiv& \frac{\sigma_{{1}/{2}}-\sigma_{{3}/{2}}}
 {\sigma_{{1}/{2}}+\sigma_{{3}/{2}}} ~,
\end{eqnarray}
where $\sigma_{1/2(3/2)}$ is the total virtual photo-absorption 
cross section for the nucleon with a projection of $1/2(3/2)$ 
for the total spin along the direction of photon momentum.

$A_2$ is a virtual photon asymmetry given by
\begin{eqnarray} A_2(x,Q^2) &\equiv&
\frac{2\sigma_{LT}}{\sigma_{{1}/{2}}+\sigma_{{3}/{2}}} ~.
\end{eqnarray}
where $\sigma_{LT}$ describes the interference between 
transverse and longitudinal virtual photon-nucleon amplitudes.
Because of the positivity limit, $A_2$ is usually small in the
DIS region and it has an upper bound given by~\cite{theory:a2soffer}
\begin{eqnarray} 
  A_2(x,Q^2)\leqslant \sqrt{\frac{R}{2}\Big[1+A_1(x,Q^2)\Big]} ~.
\label{equ:a2soffer}
\end{eqnarray} 

These two virtual photon asymmetries, depending in general on
$x$ and $Q^2$, are related to the nucleon structure 
functions $g_1(x,Q^2)$, $g_2(x,Q^2)$ and $F_1(x,Q^2)$ via 
\begin{eqnarray} 
 A_1(x,Q^2) &=&\frac{g_1(x,Q^2)-\gamma^2 g_2(x,Q^2)}{F_1(x,Q^2)} ~
  \label{equ:a1=g1f1}
\end{eqnarray}
and
\begin{eqnarray} 
 A_2(x,Q^2) &=&\frac{\gamma\Big[g_1(x,Q^2)+g_2(x,Q^2)\Big]}
  {F_1(x,Q^2)}\label{equ:a2=g1f1} ~.
\end{eqnarray} 
At high $Q^2$, one has $\gamma^2\ll 1$ and
\begin{eqnarray} 
 A_1(x,Q^2) &\approx&\frac{g_1(x,Q^2)}{F_1(x,Q^2)} ~.\label{equ:a1=g1f1-simplified}
\end{eqnarray} 

In QCD the asymmetry $A_1$ is expected to have less $Q^2$-dependence than 
the structure functions themselves because of the similar leading order 
$Q^2$-evolution behavior of $g_1(x,Q^2)$ and $F_1(x,Q^2)$.
Existing data on the proton and the neutron asymmetries $A_1^p$ and $A_1^n$ 
indeed show little $Q^2$-dependence~\cite{data:g1pg1n-e155}. 

\subsection{Electron Asymmetries}\label{ch1:easym}
In an inclusive experiment covering a large range of excitation energies
the virtual photon momentum direction changes frequently, and
it is usually more practical to align the target 
spin longitudinally or transversely to the incident electron direction
than to the momentum of the virtual photon.
The virtual photon asymmetries can be related to the 
measured electron asymmetries through polarization factors, kinematic
variables and the ratio $R$ defined in Section~\ref{ch1:unpolstrf}.  
The longitudinal electron asymmetry is defined by~\cite{exp:e080}
\begin{eqnarray} 
 A_\parallel &\equiv& \frac{\sigma_{\downarrow\Uparrow}
      -\sigma_{\uparrow\Uparrow}}
  {\sigma_{\downarrow\Uparrow}+\sigma_{\uparrow\Uparrow}} \nonumber\\
  &=& \frac{(1-\epsilon)M^3}{(1-\epsilon R)\nu F_1}
 \Big[(E+E^\prime\cos\theta)g_1-\frac{Q^2}{\nu} g_2\Big] ~, \label{equ:Apardef} 
\end{eqnarray} 
where $\sigma_{\da\Ua}$($\sigma_{\ua\Ua}$) is the cross section 
of scattering off a longitudinally polarized target, with 
the incident electron spin aligned anti-parallel (parallel) 
to the target spin, and $\epsilon$ is the magnitude of the virtual 
photon's longitudinal polarization:
\begin{eqnarray}
 \epsilon &=& \Big[1+2(1+1/\gamma^2)\tan^2(\theta/2) \Big]^{-1} ~.
 \label{equ:def-epsilon}
\end{eqnarray}
Similarly the transverse electron asymmetry 
is defined for a target polarized perpendicular to the beam 
direction as
\begin{eqnarray} 
 A_\perp &\equiv& \frac{\sigma_{\downarrow\Rightarrow}
   -\sigma_{\uparrow\Rightarrow}}
   {\sigma_{\downarrow\Rightarrow}+\sigma_{\uparrow\Rightarrow}} 
  \nonumber\\
 &=& \frac{(1-\epsilon)E^\prime M^3}{(1-\epsilon R)\nu F_1}
     \Big[g_1+\frac{2E}{\nu}g_2\Big]\cos\theta ~,~ \label{equ:Aperpdef}
\end{eqnarray} 
where $\sigma_{\da\Ra}$($\sigma_{\ua\Ra}$) is the cross section 
for scattering off a transversely polarized target with incident 
electron spin aligned anti-parallel (parallel) to the beam 
direction, and the scattered electrons being detected on the same 
side of the beam as that to which the target spin is pointing~\cite{A1nPRL}.
The electron asymmetries can be written in terms of $A_1$ and $A_2$ as
\begin{eqnarray} 
 A_\parallel &=& D(A_1+\eta A_2) \label{equ:apap1}
\end{eqnarray} 
and
\begin{eqnarray} 
 A_\perp &=& d(A_2-\xi A_1) ~, \label{equ:apap2}
\end{eqnarray} 
where the virtual photon polarization factor is given by
\begin{eqnarray}
  D&=&\frac{1-(1-y)\epsilon}{1+\epsilon R}~,
\end{eqnarray}
with $y\equiv\nu/E$ the fractional energy loss of the incident electron.
The remaining kinematic variables are given by
\begin{eqnarray}
 \eta &=& (\epsilon\sqrt{Q^2})/(E-E^\prime\epsilon)~,\\
 \xi &=& \eta(1+\epsilon)/(2\epsilon)~~~~\q{and}\\
 d &=& D\sqrt{2\epsilon/(1+\epsilon)} ~.
\end{eqnarray}

\subsection{Extracting Polarized Structure Functions from 
 Asymmetries}\label{ch1:asym2strf}

From Eq.~(\ref{equ:apap1}) and~(\ref{equ:apap2}) the virtual 
photon asymmetries $A_1$ and $A_2$ can be extracted from measured 
electron asymmetries as
\begin{eqnarray}
  A_1 &=& \frac{1}{D(1+\eta\xi)}A_{\parallel}-\frac{\eta}
  {d(1+\eta\xi)}A_{\perp} \label{equ:a2a1}
\end{eqnarray}
and
\begin{eqnarray}
  A_2 &=& \frac{\xi}{D(1+\eta\xi)}A_{\parallel}
   +\frac{1}{d(1+\eta\xi)}A_{\perp} ~.\label{equ:a2a2}
\end{eqnarray}
If the unpolarized structure functions $F_1(x,Q^2)$ and $R(x,Q^2)$ are 
known, then the polarized structure functions can be extracted from measured 
asymmetries $A_\parallel$ and $A_\perp$ as~\cite{theory:disreview}
\begin{eqnarray}
  g_1(x,Q^2) &=& \frac{F_1(x,Q^2)}{D^\prime}
  \Big[{A_\parallel+A_\perp\tan(\theta/2)}\Big] \label{equ:a2g1}
\end{eqnarray} 
and
\begin{eqnarray}
  g_2(x,Q^2) &=& \frac{yF_1(x,Q^2)}{2D^\prime\sin\theta}\nonumber\\
   &&\times \Big[{A_\perp\frac{E+E^\prime\cos\theta}
   {E^\prime}-A_\parallel\sin\theta }\Big] ~,
   \label{equ:a2g2}
\end{eqnarray}
with $D^\prime$ given by
\begin{eqnarray}
 D^\prime&=& \frac{(1-\epsilon)(2-y)}{y\big[1+\epsilon R(x,Q^2)\big]} ~.
\end{eqnarray}

\section{Formalism for e-$^3$He Elastic Scattering}\label{app:el}

The cross section for electron elastic scattering off an unpolarized 
$^3$He target can be written as
\begin{eqnarray}
 && \Big(\frac{\q{d}\sigma}{\q{d}\Omega}\Big)^{u}
    = \frac{\sigma_{Mott}}{1-\tau}
  \Bigl\{\frac{Q^2}{\vert\vec{q}\vert^2}F_c^2(Q) \nonumber\\
   &&~~~~~+\frac{\mu_{^{3}\q{He}}^2 Q^2}{2M^2}\big[\frac{1}{2}\frac{Q^2}
    {\vert\vec{q}\vert^2}-\tan^2{(\theta/2)}\big]F_m^2(Q)\Bigr\}
 \label{equ:elxsec4}
\end{eqnarray}
where $\tau\equiv Q^2/(4M_t^2)=\nu/(2M_t)$ is the recoil factor,
$M_t$ is the target ($^3$He) mass,
$Q^2$ is calculated from Eq.~(\ref{equ:qmu2}),
$\vec q$ is the three momentum transfer,
$\mu_{^{3}\q{He}}$ is the $^3$He magnetic moment, 
and $F_c$ and $F_m$ are the $^3$He charge and magnetic form factors,
which have been measured to a good precision~\cite{ana:he3ff}.
The Mott cross section $\sigma_{Mott}$ for a target of charge $Z$ 
can be written as
\begin{eqnarray}
 \sigma_{Mott}&\equiv&\Big(\frac{\q{d}\sigma}{\q{d}\Omega}\Big)_{Mott} = 
	\frac{Z^2\alpha^2\cos^2(\theta/2)}{4E^2\sin^4(\theta/2)}\frac{E^\prime}{E}~.
\end{eqnarray}
with $E^\prime$ the energy of the outgoing electrons:
\begin{eqnarray}
 E^\prime &=& \frac{E}{1+\frac{2E}{M_T}\sin^2\frac{\theta}{2}}~.
\end{eqnarray}

The elastic cross section for a polarized target can be written 
as~\cite{theory:he3elasym}
\begin{eqnarray}
   \Big(\frac{\q{d}\sigma}{\q{d}\Omega}\Big)^h &=&
	\Big(\frac{\q{d}\sigma}{\q{d}\Omega}\Big)^{u}
	 + h\Delta(\theta^*,\phi^*,E,\theta,Q^2)~,~~\label{equ:elpolxsec}
\end{eqnarray}
where $h$ is the helicity of the incident electron beam, 
$\Delta(\theta^*,\phi^*,E,\theta,Q^2)$ describes the helicity-dependent
cross section, $\theta^*$ 
is the polar angle and $\phi^*$ is the azimuthal angle of the nucleon spin 
\begin{figure}[htp]
 \begin{center} 
 \includegraphics[angle=0, scale=0.7]{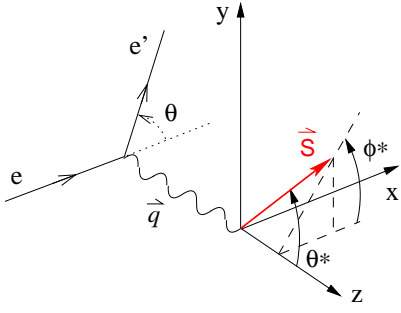}
\end{center}
 \caption{(Color online) Polar and azimuthal angles of the target spin.}
 \label{fig:elkine}
\end{figure}
direction, as shown in Fig.~\ref{fig:elkine}.  We write them 
explicitly for a target with spin parallel to the beam direction 
as~\cite{theory:he3elasym}
\begin{eqnarray}
 \cos{\theta^*} &=& (E-E^\prime\cos{\theta})/\vert\vec q\vert \\
 \phi^* &=& 0~.
\end{eqnarray}
The helicity-dependent part of the cross section can be written as
\begin{eqnarray}
   \Big(\frac{\q{d}\sigma}{\q{d}\Omega}\Big)^{h=+1}
     \hspace*{-0.3cm}-\Big(\frac{\q{d}\sigma}{\q{d}\Omega}\Big)^{h=-1}
     \hspace*{-0.3cm} &=&
     - \sigma_{Mott}\Big( V_{T^\prime}R_{T^\prime}(Q^2)\cos\theta^* 
      \nonumber \\
     && \hspace*{-1.4cm}
        + V_{TL^\prime}R_{TL^\prime}(Q^2)\sin\theta^*\cos\phi^* \Big)~,
   \label{equ:Apar-el-dxsec}
\end{eqnarray}
with kinematic factors
\begin{eqnarray}
  V_{T^\prime} &\equiv& \tan{\frac{\theta}{2}}\sqrt{\frac{Q^2}
   {\vert\vec{q}\vert^2}+\tan^2{\frac{\theta}{2}}}
\end{eqnarray}
and
\begin{eqnarray}
  V_{TL^\prime}&\equiv& -\frac{Q^2}{\sqrt{2}\vert\vec{q}\vert^2}
   \tan{\frac{\theta}{2}}~.
\end{eqnarray}
$R_{T^\prime}$, $R_{TL^\prime}$ can be related to the $^3$He
form factors $F_c$, $F_m$ as:
\begin{eqnarray}
 R_{T^\prime} &=& \frac{2\tau E^\prime}{E}(\mu_{^3\q{He}} F_m)^2
\end{eqnarray}
and
\begin{eqnarray}
 R_{TL^\prime} &=& -\frac{2\sqrt{2\tau(1+\tau)}E^\prime}{E}(ZF_c)(\mu_{^3\q{He}} F_m)~,
\end{eqnarray}
The elastic asymmetry, defined by
\begin{eqnarray} \label{equ:he3elasymdef}
  && A^{el}_{\parallel} \equiv
    \frac{\big(\frac{\q{d}\sigma}{\q{d}\Omega\q{d}E^\prime}\big)^{h=+1}
	-\big(\frac{\q{d}\sigma}{\q{d}\Omega\q{d}E^\prime}\big)^{h=-1}}
	{\big(\frac{\q{d}\sigma}{\q{d}\Omega\q{d}E^\prime}\big)^{h=+1}
	+\big(\frac{\q{d}\sigma}{\q{d}\Omega\q{d}E^\prime}\big)^{h=-1}}~,
\end{eqnarray}
can therefore be calculated from Eq.~(\ref{equ:elxsec4}) 
and~(\ref{equ:Apar-el-dxsec}) as
\begin{eqnarray} \label{equ:he3elasymcal}
  && A^{el}_{\parallel} =  -(1-\tau) \nonumber\\
  && \hspace*{0.2cm}\cdot\frac{\big({V_{T^\prime}R_{T^\prime}\cos\theta^*
	    +V_{TL^\prime}R_{TL^\prime}\sin\theta^*\cos\phi^*}\big)}
	{\Bigl\{\frac{Q^2}{\vert\vec{q}\vert^2}F_c^2
	+\frac{\mu^2 Q^2}{2M^2}\Big(\frac{1}{2}\frac{Q^2}{\vert\vec{q}\vert^2}
	-\tan^2{\frac{\theta}{2}}\Big)F_m^2\Bigr\}}~~~
\end{eqnarray}


\vspace*{1.4cm}
\centerline{\bf{ACKNOWLEDGMENTS}}

We would like to thank the personnel of Jefferson Lab for their efforts which 
resulted in the successful completion of the experiment.
We thank S.~J.~Brodsky, L.~Gamberg, 
N.~Isgur, X.~Ji, E.~Leader, W.~Melnitchouk, 
D.~Stamenov, J.~Soffer, M.~Strikman, A.~Thomas, M.~Wakamatsu, H.~Weigel 
and their collaborators for theoretical support and helpful discussions. 
This work was supported by the Department of Energy (DOE), 
the National Science Foundation,
the Italian Istituto Nazionale di Fisica Nucleare,
the French Institut National de Physique Nucl\'{e}aire et de Physique des 
Particules,
the French Commissariat \`{a} l'\'{E}nergie Atomique
and the Jeffress Memorial Trust. 
The Southeastern Universities Research Association operates the Thomas 
Jefferson National Accelerator Facility for the DOE under contract 
DE-AC05-84ER40150.\\




\end{document}